# Tunable structure-activity correlations of molybdenum dichalcogenides (MoX$_2$; X= S, Se, Te) electrocatalysts via hydrothermal methods: insight into optimizing the electrocatalytic performance for hydrogen generation

Zhexu Xi

*Bristol Centre for Functional Nanomaterials, University of Bristol, Bristol, UK*

## Abstract

Hydrogen Evolution Reaction (HER) has always gained wide attention as one of the eco-friendly and sustainable pathway for efficient hydrogen generation and storage; also, two-dimensional molybdenum dichalcogenide (MoX$_2$, where X stands for S, Se, Te) layers have emerged as a class of quasi-ideal electrocatalysts because of their large surface area, rich reserves and outstanding conductivity. However, besides greater HER activity, the maturity and diversity of modification strategies result in more puzzling relationship between electrocatalytic mechanisms and the corresponding practical performance. In this article, based on a comprehensive review of fundamentals, principles and interconnected similarities of the MoX$_2$ family, we focus on the structure-activity correlation of layered MoX$_2$ for HER enhancement via hydrothermal synthesis. This method is summarized from different experimental systems to efficiently modulate the crystal structure and surface for boosted HER activity. Here, with the adjustment of three key experimental parameters: the categories of MoX$_2$, reaction temperature and the molar amount of added reactants, the optimum HER performance can be obtained at the best conditions (MoSe$_2$ species, 180°C and a vast ratio of the reductant or metal precursor), and more microscopically, a controlled structure-activity relationship can be inducted. This summary may pave a new path for the controllable synthesis and modification of MX$_2$-based catalyst materials.

**Key words:** *molybdenum dichalcogenides; structure-activity analysis; tunable; phase transition; active sites; hydrogen evolution; hydrothermal*

# Content

## 1. Introduction

As the environmental pollution and energy crisis become increasingly severe, hydrogen, owing to its tiptop energy density, renewability, high purity and zero-polluting combustion byproduct (water), has received greater attention as an ideal energy carrier to reduce the dependence on traditional fossil energy[1-3]. Over various hydrogen generation pathways (in **Fig. 1**), water splitting via electrochemical approaches has been regarded as a low-cost, eco-friendly and sustainable industrial pathway for high-efficiency hydrogen conversion and storage[4,5]. So far, numerous experimental studies about high-speed and efficient hydrogen evolution have been gradually categorized into two classes: 1) identifying HER mechanisms in pursuit of more strategies for accelerating HER reaction rates, especially at a wide range of pH containing neutral and alkaline electrolyte environments (theoretically)[6-12]; 2) the discovery and design of new kinds of durable and high-activity HER electrocatalysts (experimentally)[13-18].

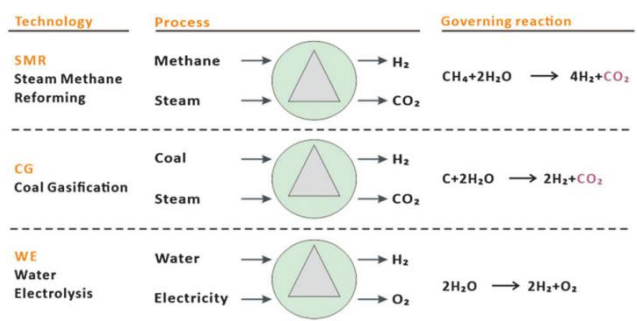

**Fig.1** Three common industrial pathways for hydrogen generation[5].

Considering the class 1), the key to understanding the HER mechanism is to explore the inherent relationship between the microscopic viewpoint of intermediate adsorbed states (including intermediate species and the triggered activation and adsorption energy change) and the macroscopic reaction rates[19]. Although the perplexing principle of the HER process in different pH conditions (mainly referring to acidic, neutral and alkaline conditions) is still under heated debate, especially considering which factor plays a predominant role including the source of proton donors[6,7], the interfacial H*-M (hydrogen-metal) band intensity with the changed activation barriers[8,9], the availability of surface sites and electron trapping states[10], $H_{upd}$[11], pzfc (the potential of zero free charge) with the changed reorganizational energy[12], there is a common consensus based on the competing
3

relationship between the extra water dissociation and activation step and the hydrogen adsorption/desorption step. Specifically, from the perspective of catalyst design, several feasible strategies should be implemented to accomplish two goals (as **Fig.2** depicts[20]): improving the reaction thermodynamics by lessening the activation barrier from dissociated water molecules (e.g. creating more oxophilic sites); promoting the reaction kinetics by tuning the H*-M interactions (e.g. modulating the electronic structures)[20,21]. Accordingly, no matter what the respective value of two goals are in HER, more micro-to-macro relationship can be established between the HER-related principles and the apparent HER activity by taking theoretically well-defined surface structures and electronic band levels of a certain electrocatalyst into account.

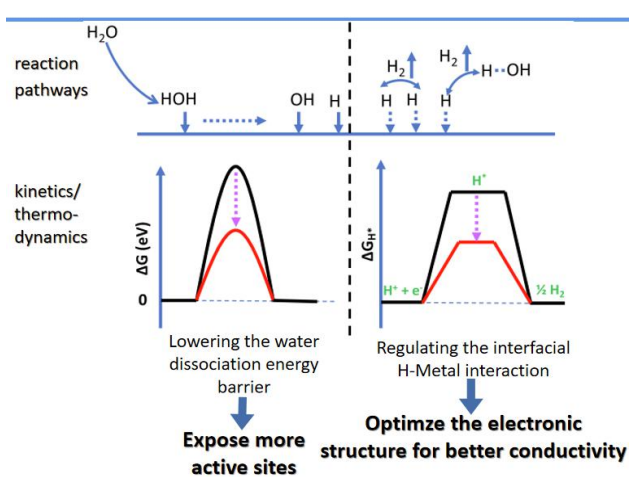

**Fig.2** Schematic illustration of two goals and approaches as examples linked with two competing steps as the possible HER mechanisms. Here, the idea is to combine possible mechanisms with the practical design of a well-designed electrocatalyst with a well-defined surface and a well-tuned electronic structure from the thermodynamic and the kinetic perspectives.

Another class of research entails the real-world design of a certain kind of high-activity electrocatalysts. Although noble metals with their compounds, especially platinum (Pt), exhibit the optimum HER activity according to the Sabatier principles[22], their rare reserves and exorbitant prices largely restricts the large-scale hydrogen production. With a comparably low overpotential, a low Tafel slope and a moderate $\Delta G_{H*}$ (not too big or too small) to Pt, various materials have adequate potentialities to replace the Pt-based HER catalysts, including chalcogenides[13,14], oxides[15], phosphides[16], nitrides[17] and carbides[18] ranging from bulk to nanoscale. Fully considering important structural or physical properties like surface area, crystallinity, porosity, thickness, electron conductivity and layered assemblies, molybdenum dichalcogenides ($MoX_2$) have superior activity and long-term durability to defeat other structured catalyst materials[23-26]. Accordingly, the suitable choice of $MoX_2$ help govern and regulate the apparent reactivity and kinetics of HER by designing a practically high-performance electrocatalyst with controlled surfaces and morphologies from a theoretically well-defined catalyst surface based on the HER principles.

Consequently, our work aims to provide a comprehensive structure-activity analysis of $MoX_2$-based electrocatalysts to present a clear mapping between the sluggish-rate-related HER energetics of two intermediate thermodynamic states (produced by two competing steps: the extra water dissociation step and hydrogen adsorption with interfacial H*-M interactions, as shown in **Fig. 3 (b)**[27,28]) and the practical design of a high-activity electrocatalyst. Based on the aforementioned correlations among hydrogen generation and two classes of viewpoints (simplified in **Fig. 3 (c)**), the commonly critical issue is nanosurfaces and nanostructures design, not only for boosted electrocatalytic performance and clearer understanding of alkaline HER, but also for better tuned HER reaction kinetics.

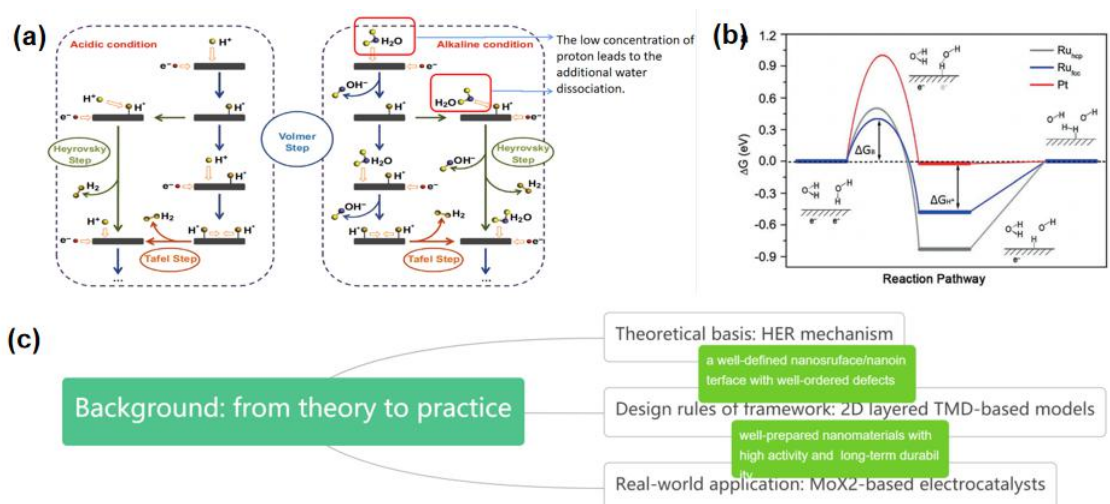

**Fig. 3 (a)** scheme showing the different HER mechanisms in acidic and alkaline medium which identifies the reason for the sluggish rate of alkaline HER[29]: the additional water dissociation step with formation of the extra activation barrier ascribing to the lack of proton sources (H*)[27]; **(b)** Gibbs Free Energy diagram of alkaline HER on different electrode surfaces showing the energetics between different thermodynamic states during the water dissociation and hydrogen adsorption steps[28]; **(c)** schematic diagram of the relationship between two classes of viewpoints from a theoretical model to a practically high-performance electrocatalyst.

In our work, we attach importance to the structure-activity correlation of MoX$_2$-based electrocatalysts for enhanced HER activity as well as kinetics. Specifically, considering fully the potential approaches originating from the two competing pathways in HER, especially the extra water dissociation step in neutral/alkaline media, the extrinsic physical/morphological structures and intrinsic electronic band structures need to be taken into account to signify their roles in promoting the reaction rates. As a result, a hydrothermal approach, as one of the convenient and straight-forward techniques in tunable synthesis of layered transitional metal dichalcogenides (TMDs), can be applied widely to better control the structure-based features (e.g. size, shape, phase, morphology and composition) of products by adjusting several experimental parameters like kinds and amount of precursors, reductants and solvents, temperature and reaction time[30-32]. In addition, the altered parameters under facile operational procedures and mild conditions make access to a clearer, better tuned structure-activity relationship, such as regulating the exposure of surface active sites[31].

Here, a more comprehensive structure-activity correlation in the MoX$_2$-based (including sulfides, selenides and tellurides) HER experimental system is studied in two key aspects: 1) Can the unique features (especially physical, structural and electronic properties) of the family of molybdenum chalcogenide compounds contribute to more revelation of HER catalytic performance? 2) By employing different parameters in hydrothermal procedures, can microscopically structural factors contribute to more precise tuning of HER activity and kinetics? In the first aspect, the fundamentals of the layered MoX$_2$ family is summarized to signify their inherent superiority in structure-related features, containing the overall excellent properties as entities, such as the roles of active sites and crystal phases, and the property changes among the individuals. In the second aspect, following the guidelines of optimizing water dissociation and H* adsorption steps, three key parameters (the categories of MoX$_2$, reaction temperature and the molar amount of added reagents) is comprehensively analyzed to bring about the controlled morphology and thereby the prompted HER activity. Here, according to the design rules of a promising electrocatalysts (typically, referring to a low Tafel slope, a low overpotential or/and a high exchange current density) elucidated from the measured values in electrochemical tests[33,34], hydrothermal procedures generate a more tunable electrocatalytic HER system with a more interpretable structure-activity tuning mechanism.

## 2. Fundamentals and principles of layered molybdenum dichalcogenides: a reliable and promising electrocatalyst family

### 2.1 Overview

As **Fig.2** depicts, owing to the extra water adsorption and dissociation on the catalyst surface, the low concentration of proton sources (hydrate) in neutral or alkaline medium severely constrain the HER activity considering both thermodynamic (activation barriers and overpotential, as illustrated in **Fig.3(b)**) and kinetic parameters (exchange current density and Tafel slope)[35]. Although the increasing electrolyte pH inevitably results in a more sophisticated mechanism with the four-electron HER process, numerous computational research has emphasized three corresponding indicators for the rational design of advanced HER electrocatalysts[36-38].

The adsorption-free energy ($\Delta G_{H^*}$) is the most prevalent to describe the ability of an electrocatalyst to initiate the reaction. It demands the hydrogen binding strength at the catalyst-electrolyte interface to reach an appropriate value (close to zero) so that the hydrogen adsorption and desorption processes could reach an optimal balance[35,39,40]; however, apparently, it neglects the additional participation of activated water molecules. The other descriptors are water adsorption energy ($E_{ad}$) and the activation energy of water dissociation ($E_{ac}$). The lower values of the two descriptors indicate the faster kinetics of adsorbed and dissociated water molecules, thereby prompting the overall reaction. Accordingly, as the comprehensive analysis of the HER energetics from both thermodynamic and kinetic perspectives, two corresponding strategies-the surface enrichment of active sites and increase of intrinsic activity for optimized electronic states-emerges based on the better understanding of these interconnected indicators, in agreement with the schematic revelation of **Fig.2**.

As a competitive alternative to Pt, 2D layered $MoX_2$ materials have their unique characteristics and the corresponding modification techniques for the rational design of a type of durable, highly efficient and cost-effective HER electrocatalysts. First, their comparably high activity and stability to noble-metal-based catalysts, as well as low cost, easy accessibility and non-scarcity, significantly facilitate their large-scale application, meeting the demands for long-term advancements[41]. Second, following the design rules of HER electrocatalysts, in order to create a well-defined nanostructure/nanosurface with well-ordered defects in guarantee of lowering the water dissociation barriers and governing the hydrogen adsorption kinetics, versatile modification strategies have been put forward to optimize the electronic structures and modulate the structural morphologies of $MoX_2$ materials like surface area, interlayer thickness, edges, facets, phases, crystallinity, alignment and so on[30-32,34,42-46]; especially, a huge surface area and various crystal phases of molybdenum chalcogenides are two critical and well investigated structural features in promoting the HER electrocatalytic performance, which will be introduced in detail in **2.2** and **2.3** for the identification of their roles.

For clearer structure-function-methodology regimes, a system of generalized guidelines have been summarized based on diversified strategies for the rational design of $MoX_2$-based catalysts: (1) increasing the densities of active sites by larger exposure of surfaces and edges, or creating new sites inside layers; (2) activating the inherently inert basal plane for optimized H* adsorption strength and modulating the intrinsic activity; (3) improving the electron conductivity by tuning electronic band structure[47,48].

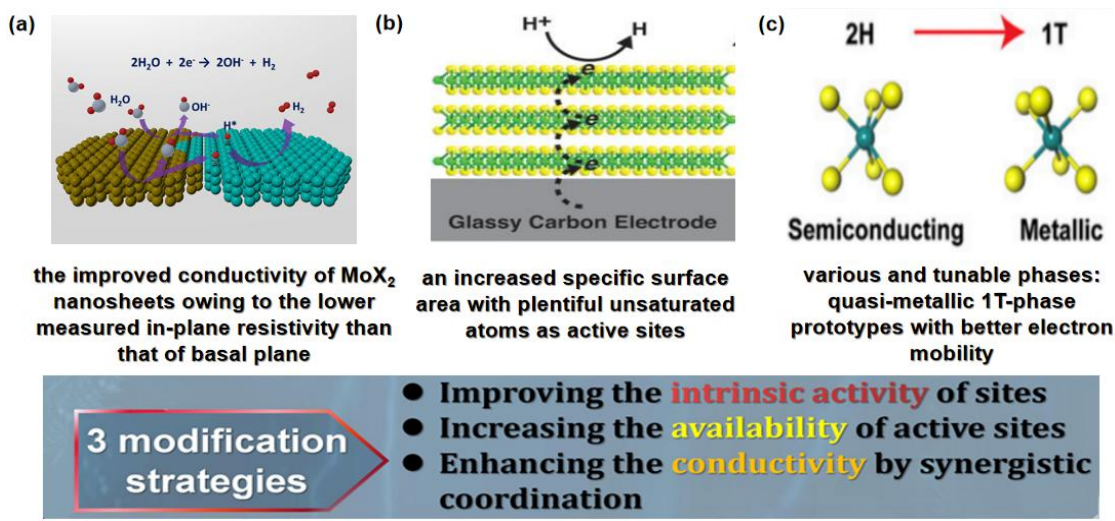

**Fig. 4** Schematic illustration of how the emerging modification strategies for MoX$_2$ nanostructures design corresponds to the characteristics of these materials[49]. The general idea of these structure-guided strategies refers to three aspects: density of unsaturated active sites, activity of in-plane and edge sites; electron band structure engineering.

## 2.2 The role of active sites in MoX$_2$

Nanoengineering provides unprecedented opportunities for 2D MoX$_2$ layers to exhibit evidently enhanced electrocatalytic performance as well as more higher atomic utilization efficiency in the structure of nanosheets (NSs)/nanoplates[50]. Compared with the bulk or 0D counterparts, one of the underlying reasons is their ultrahigh specific surface area with plenty of coordinately unsaturated atoms serving as surface active centers. Although there is an inevitable trend that the as-synthesized 2D nanostructures are likely to thermodynamically generate 0D nanoparticles (NPs) to keep the molecular surface energy at a lower level, diversified methods can be conducted in a wider way for larger exposure of surface sites and less degree of 2D-0D transformation[50,51]. Furthermore, the ultrathin thickness of 2D layers results in a bigger surface area to volume ratio, thereby having more possibilities to expose embedded active sites, especially edge sites.

Accordingly, the large area of layered MoX$_2$ provides enriched active sites on the surface, including edge sites and in-plane sites. For edge sites, nanostructures design, such as vertical arrays[52,53] and mesoporous structures with higher porosity[54,55], is widely used to create or expose more sites to lower the activation energy and tune the thermodynamics in HER as suggested in **Fig.2**. Kong et al.[52] constructed the vertically aligned MoS$_2$ and MoSe$_2$ layers during the formation of thin films, maximally exposing the catalytically active sites on edges as shown in their idealized structures of **Fig. 5(a)**. Correspondingly, the maximal active sites brings the optimal HER performance with strong activity (according to the measured values of Tafel slope and exchange current density of MoS$_2$ and MoSe$_2$ samples in **Fig. 5(b)**) and stability (**Fig. 5(c,d)**). Likewise, in the work of Lukowski et al.[53], with the basal plane-based vertical arrays to the substrate, the edge sites of the as-synthesized MoS$_2$ single-crystal nanobelts realize their largest surface exposure, thereby leading to the evidently boosted hydrogen evolution efficiency.

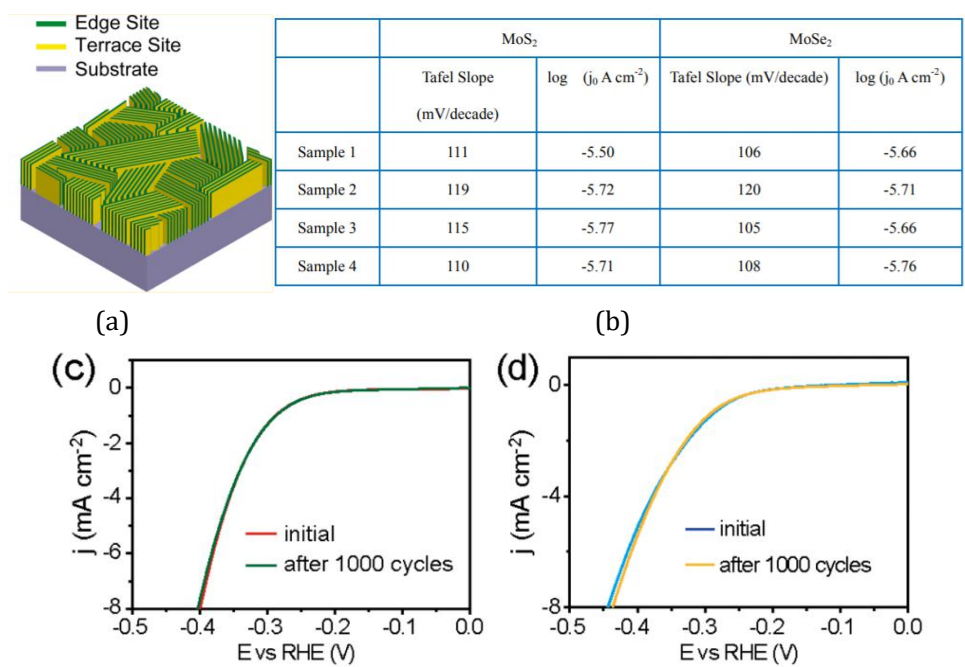

**Fig.5 (a)** Idealized structure of edge-terminated molybdenum chalcogenide films with the layers aligned perpendicular to the substrate, maximally exposing the edges of the layers; **(b)** the chart derived from the electrochemical measurement. The average exchange current density are 2.2×10$^{-6}$ A/cm$^2$ for MoS$_2$ and 2.0 ×10$^{-6}$ A/cm$^2$ for MoSe$_2$; HER electrocatalyst stability test for **(c)** MoS$_2$ and **(d)** MoSe$_2$ in which negligible HER currents are lost after 1000 cycles in the cathodic potentials windows[52].

Based on the above special types of nanoengineered structures, the key idea is to increase the edge length to maximize the exposure of edge sites, simultaneously further demonstrating the early discovery that the catalytic performance of MoS$_2$ nanoplatelets is positively correlated with the length of edge state[56]. For the same purpose, thinning the thickness of MoX$_2$ layers by reducing the layer numbers is also an effective strategy. For instance, exfoliation from the bulk-state materials is a useful and direct method to obtain monolayer or few-layer products, but because of the complex procedures in liquid, it's difficult to generate the products with desired properties like controlled morphologies and interfaces, and inuniform size. The exfoliated molybdenum chalcogenides present excellent HER performance with a Tafel slope of 94.91 mV dec$^{-1}$ and onset potential of approximately 100 mV (at 1 mA/cm$^2$)[57]. Further, in order to form a high-yield, ultra-thin nanosheet, different ion intercalation process can be added in liquid-assisted exfoliation for more drastically enhanced catalytic efficiency[58]. The intercalation of metal ions into the interlayer region of MoX$_2$ induce an interlayer distancing change and lead to more exposed active edge sites and X-vacancies on surfaces for increased electrochemical active surface area (ECSA); besides, the intercalation modulate the local electronic states, thereby optimizing the hydrogen adsorption with the altered $\Delta G_{H*}$. In Daneil et al's work[59], various cations including Na$^+$, Ca$^{2+}$, Ni$^{2+}$, and Co$^{2+}$ were inserted into the 1T-MoS$_2$ interlayers for HER performance test (**Fig. 6**). Compared with the result from the control group (intercalant-free MoS$_2$), the $\Delta G_{H*}$ values of the intercalated structures show a marked descent, presenting an enhanced activity with various extent of improvement. Particularly, the $\Delta G_{H*}$ of Na$^+$ intercalated MoS$_2$ is the closest to zero, revealing the supreme HER activity.

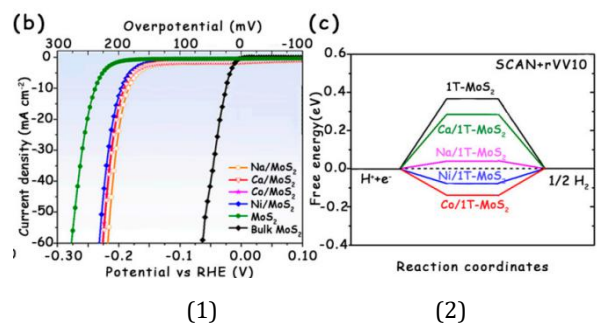

**Fig. 6 (1)** Current density-overpotential plot (polarization curve, after iR correction) of bulk 2H-, 1T-MoS$_2$ and MoS$_2$ intercalated with different metal cations, indicating improved electrocatalytic activity of the ion intercalated products; **(2)** schematic comparison in free energy change of hydrogen adsorption of the products, where the general idea is the closer to zero free energy is, the more comparative activity the material possesses to Pt[59].

Although the entire catalytic activity of layered molybdenum catalyst materials depends mainly on the availability of active sites at edges rather than area coverage[56], the vast majority of sites exist on the basal plane. Originally, the basal plane is always regarded to be inherently inert, so introducing additional active sites is of great importance to boost the intrinsic in-plane activity for optimized HER behavior. As a typically comprehensive computational work, Lin et al.[60] surveyed the in-plane activity of defected MoA$_2$ (A=O, S, Se), discovering that the defect states on the surface can capture the hydrogen atoms, further modulating the local band structure and tuning the adsorption energy to promote the intrinsic activity of basal plane. To investigate more clearly the role of different defects and surface states in HER, Wang et al.[61] identified that chalcogenide-induced vacancies (V$_X$, X=S, Se, Te), MoX$_3$ vacancies (V$_{MoX_3}$), 4|8a dislocation-induced fold rings, Mo-Mo bond grain boundaries, X bridge and MoX$_2$ point defects can accelerate the Heyrovsky and Tafel step with the tuned electronic structures and the optimized interfacial hydrogen adsorption, thereby promoting the HER kinetics. Based on similar theoretical guidance to the above, Vasu et al.[62] investigated the influence of doped Ru atoms on the MoSe$_2$-based catalytic performance by adjusting the proportion of Ru dopants. From the XRD result shown in **Fig. 7(a)** and **(c)**, the doped samples exhibit their typical Bragg reflections of MoSe$_2$ with no formation of secondary phases, indicating that the Ru dopants are not absorbed onto the Mo surface or intercalated among layers, but anchored within the basal plane as the substitutions to Mo. In detail, the XPS result in **Fig. 7(c)** highlights the intimate dependence between the electronic structure of the doped samples and the content of Ru: a noticeable binding energy upshift reveals the influence of the enhanced in-plane n-type conductivity and the tuned electronic states near the Fermi level. In addition, the electrochemical test (**Fig. 7(b)**) signifies the improved HER activity, where the largest content of Ru corresponds to the fastest charge transfer kinetics (according to the lowest resistivity from the Nyquist plot) and the optimal activity.

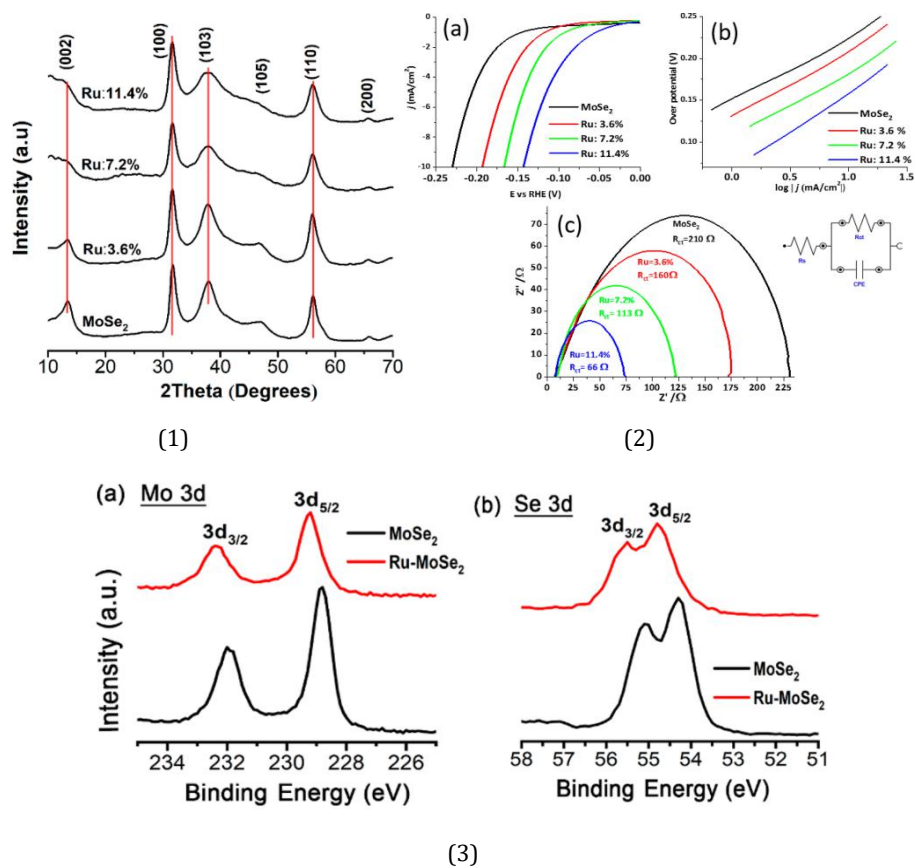

(1)

(2)

(3)

**Fig. 7 (1)** XRD pattern of pure and Ru-doped MoSe$_2$ for different contents of Ru, mainly showing a maintained similar morphology at nanoscale with no apparent new phases produced; **(2)** results of HER measurements of MoSe$_2$ and various Ru-doped MoSe$_2$ in **(a)** polarization curves, **(b)** the corresponding Tafel plots and **(c)** Nyquist plots from EIS tests at the same overpotential of 200 mV and operation cycles of 2000 r, showing the content of Pu dopants is correlated positively to the overpotential and negatively to the charge transfer resistance with the nearly unchanged, comparatively low value of Tafel slope, thereby suggesting the promoted HER activity; **(3)** XPS core-level spectra pattern of **(a)** Mo 3d and **(b)** Se 3d states of pure and 11.4% Ru-doped MoSe$_2$[62].

Generally, in-plane activation has established close links to the increasing intrinsic activity of active sites on basal plane. Specifically, in-plane sites with appropriately plentiful densities and concentrations act as oxophilic centers for facilitated hydrogen adsorption and subsequently modulated electronic structures of catalysts; conversely, excess available sites serving as electron trapping agents may in return impede the electron mobility and further deteriorate the practical performance of MoX$_2$-based HER catalyst. Furthermore, in defect and disorder engineering, versatile methods may lead to different degrees of structural imperfection and various types, numbers and distribution of defect states, indicating a hardship in well-controlled morphologies.

**2.3 The role of phase structures in MoX$_2$**

The reason for the MoX$_2$ family acting as a promising HER electrocatalyst is also closely connected with their multiple and well-tuned crystal phases. There are five polytypes in terms of the crystal structures of MoX$_2$, which are exhibited in **Fig. 8**[63]. Their intralayer bonds are covalent while the interlayer bonds are van der Waals. Different chalcogenide atoms determine various physicochemical combinations with a certain degree of similarity in MoX$_2$, thereby resulting in diverse but interconnected electronic properties[63,64]. In addition, by choosing different elements of chalcogenide, different phases in monolayer can be obtained. Among them, 1T, 2H and 3R phase types are the most common structure polymorphs, where subtle structural change may influence the properties strikingly. The 2H-phase are favorably applied as the as-prepared raw materials for design of electrocatalyst due to its thermodynamic stability based on the semiconducting characteristics; the metastable 1T- or 1T'-phase (distorted 1T-phase) with more emerged metallic properties have more potentials to contribute to the enhanced HER activity. Consequently, converting 2H to 1T/1T' phases can be a mainstream strategy to significantly improve the catalytic performance[63,65].

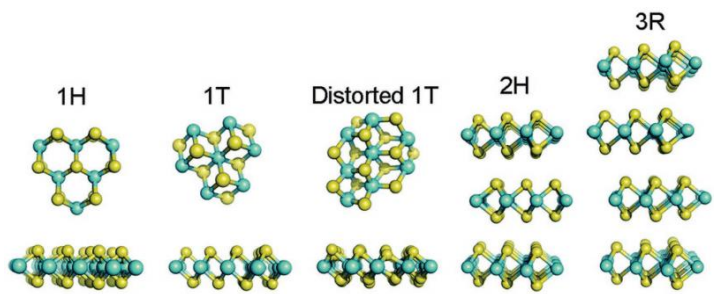

**Fig. 8** Schematic of five categories of crystal structures of molybdenum dichalcogenides, from left to right: 1H phase, 1T phase, distorted 1T (1T') phase, 2H phase and 3R phase[63].

For more feasible and tuned phase transition, numerous methods have been well investigated:

(1) alkali metal intercalation

As aforementioned, the intercalation of a sp-hybridized or d-band metal cation into the interlayer regions will lead to the formation of a new electronic state, subsequently bringing about the additional electron transfer in the proton reduction cycle (from reducing agents to MoX$_2$ catalysts). Accordingly, the stability of the pristine 2H phase will be weakened by overloaded d-electrons in Mo, thereby generating the tendency of formation of a new 1T phase[66]. A high-yield electrochemical lithium insertion method was experimented by Jiang's group[67] to demonstrate a 2H-1T phase transition of $MoSe_2$ nanosheets in the initial discharge process at 0.9 V. Specifically, they identified the corresponding $Li^+$ diffusion kinetics in layers of both $Li^+$-intercalated crystalline and amorphous samples that the transition process was in pace with the formation of tetrahedron sites as the energy lowest configuration. After that, due to the lower electron conductivity from excessive concentration of active sites, the amorphous $MoSe_2$ was demonstrated to have lower reaction rates compared with the crystalline one. Meanwhile, volume expansion and lattice distortion were further observed during the first lithiation to induce more surface defects and create a larger catalytically active surface area[67,68].

(2) strain engineering

Introducing a compressive or tensile strain can contribute to the tuning of $MoX_2$ properties. Applying strains on the structure leads to a localized lattice distortion, thereby activating the basal plane and optimizing the hydrogen adsorption. Finally, strain engineering plays a predominant role in promoting electrocatalytic activity[69]. Notably, the extent of the strain needed for a successful phase transition is different among various molybdenum chalcogenide materials, but is below the threshold for destroying the layer structure. $MoTe_2$ was calculated and experimented (with a flexibly supportive substrate, shown in **Fig. 9**) to be the best candidate to realize a strain-induced phase transition at the optimal conditions of a minimal strain of 0.3-3% at room temperature. Chemically deposited on the substrate, the distorted 1T-$MoTe_2$ samples under test at the best conditions was still verified to have a superior catalytic activity with an extraordinarily low Tafel slope of 46.3 mV/dec[63,70].

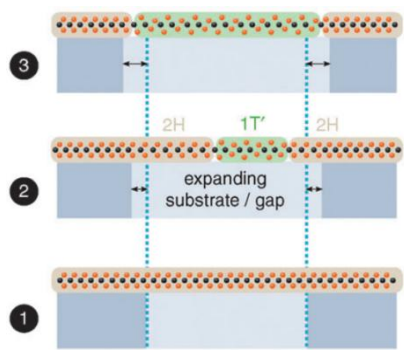

**Fig. 9** 2H to 1T phase transformation of single-layer TMDs when applying uniaxial strain. Here, the region quasi-freely suspended or on a low-friction substrate (middle) is converted[63].

Furthermore, Park et al.[71] introduced a mechanical lattice strain in the momolayer MoSe$_2$ via in-situ Se-vacancy engineering for prompted HER performance, where the modulation of Se-vacancy was conducted during the CVD-assisted growth of samples by adjusting the hydrogen gas concentration. For HER performance of vacancy-modulated samples (**Fig. 10(2-4)**), MoSe$_2$-50, which means the samples synthesized in 50% H$_2$ atmosphere, reveals an outstanding performance with a low onset potential of around 0.16 V and a comparably low Tafel slope (0.33 mV/dec) and the closest ΔG$_{H^*}$ to Pt/C reference. Among catalytic activity of the samples synthesized in five different hydrogen concentration, also considering the growth mechanism (**Fig. 10(1)**), fewer impurities (e.g. MoO$_{3-x}$ clusters) and more defect as active sites indicate better activity. More importantly, the vacancy-dependent catalytic behavior was elucidated that Se-vacancies served as hydrogen adsorption sites for accelerated Volmer step by creating new localized electronic states near the Fermi level and controlling the partial electron density, as well as lowering the activation barriers of hydrogen diffusion for Tafel reaction enhancement, as illustrated in **Fig. 10(5)**.

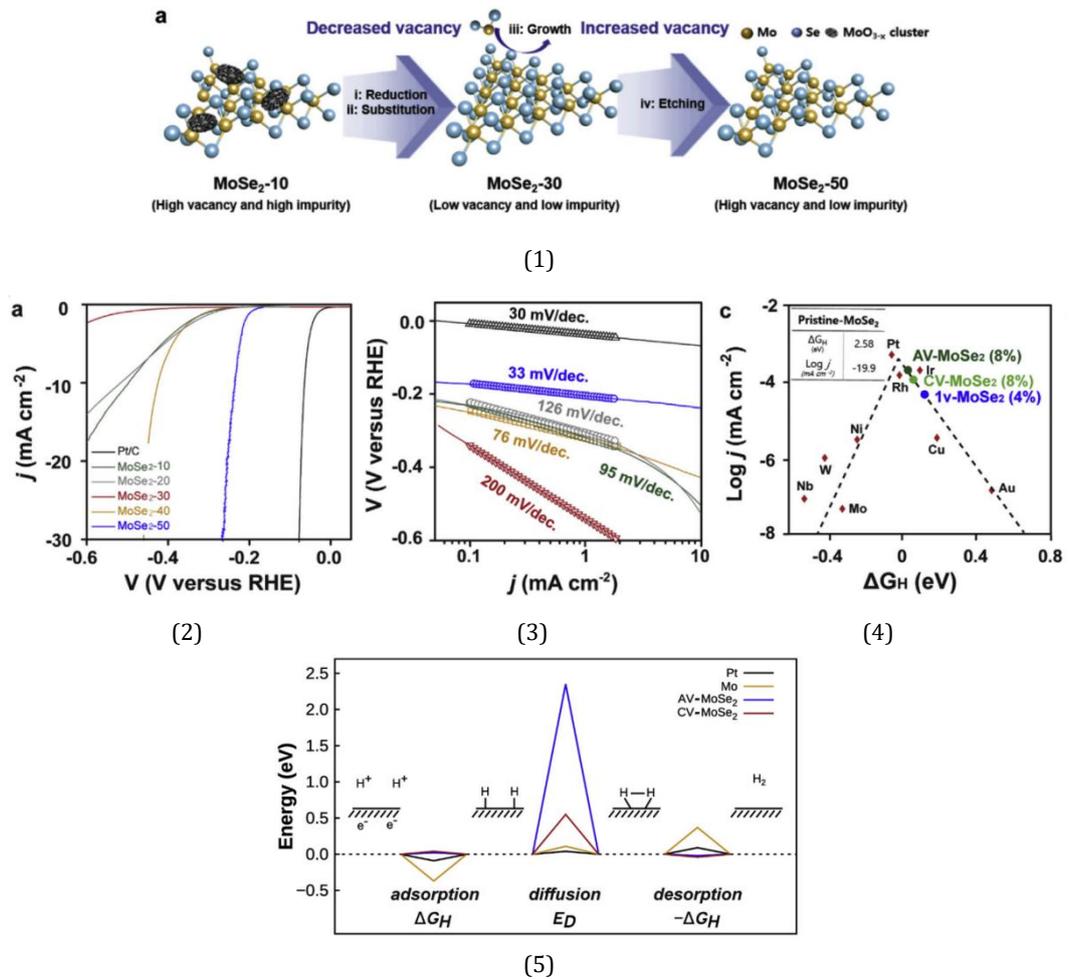

**Fig. 10 (1)** Schematic illustration of synthesis mechanism of vacancy-modulated MoSe$_2$ by H$_2$ concentration control. Here, the MoSe$_2$ samples synthesized in the H$_2$ concentration of x% are expressed in the form of MoSe$_2$-x. Notably, hydrogen plays a dual role in promoting the reduction of MoO$_3$ (decreased vacancies) and etching Se at the MoSe$_2$ lattice (increased vacancies); **(2)** polarization curves and **(3)** the corresponding Tafel plots from the HER test are used to evaluate the effect of Se-vacancy in HER activity of MoSe$_2$; **(4)** volcano plot presenting the comparison of exchange current density (log $j$) and hydrogen adsorption-free energy ($\Delta G_H$) among pristine metals and vacancy-induced MoSe$_2$, indicating that two categories have the comparable catalytic performance. The smaller inset chart shows the $\Delta G_H$ and log $j$ of pristine-MoSe$_2$ (vacancy-free); **(5)** schematic illustration of the overall energy change in the Volmer-Tafel pathway of HER including H$_3$O$^+$ adsorption, H* diffusion and H$_2$ desorption. The general idea is that the CV-MoSe$_2$ (coalesced Se-vacancies-induced MoSe$_2$) can evidently lower the hydrogen diffusion barrier for the accelerated Volmer and Tafel reactions compared with AV-MoSe$_2$ (apart)[71].

(3) template-assisted synthesis

The Park's survey indicated that besides various vacancies as active sites, the H-1T transformation could change the rate determining step (RDS) from the Volmer to the Heyrovsky reaction, accordingly regulating the interfacial H*-catalyst interaction and facilitating the HER kinetics[71,72]. A facile strategy of incorporating 2H-MoX$_2$ onto the reduced graphene oxide (RGO) template can markedly increase the percentage of 1T phase type[73-75]. RGO as a well-investigated electron donor can accelerate the electron transfer behavior with $MoX_2$, so the phase transition is promoted. For example, Wei's group[73] utilized the RGO template to modulate the 1T phase in $2H-MoS_2$ for monitoring the HER enhancement. Compared with the pristine samples, the modified samples with 50% 1T phase reveals a satisfactorily boosted HER performance with an attractive Tafel slope of 35 mV/dec, a low overpotential of 126 mV (at the exhcnage current density of 10 $mA/cm^2$) and a negligible charge transfer resistance of 12 ohm, as well as an outstanding long-term stability, as depicted in **Fig. 11**. More precisely microscopic views further indicated that 2H-1T transformation contributed to the altered electron density near the Fermi level and adjusted the formation of hydrogen intermediates by diminishing the formation energy barrier.

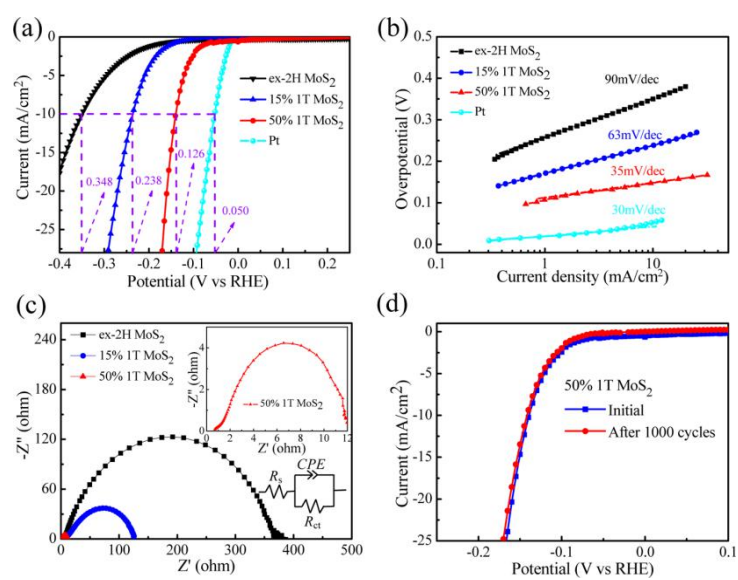

**Fig. 11** HER electrochemical measurement among commercially used Pt and 2H $MoS_2$ with different contents of 1T type: **(a)** polarization curves; **(b)** corresponding Tafel plots; **(c)** Nyquist plots derived from EIS test; **(d)** stability assessment via accelerated cyclic voltammetric test[73].

On the basis of the above feasible methods for making full and flexible use of phases for HER enhancement, the investigations of the role of 1T phase revealed the multiple underlying reason for the enhanced activity of 1T-type $MoX_2$: the synergistic effect of the facilitated interlayer electron mobility and the larger exposure of basal sites by in-plane activation, compared with the 2H phase[74,76]. Moreover, the content of 1T/2H types in the phase structure of $MoX_2$ can be further regulated for better understanding of HER tuning principles. Plentiful similar research shown the HER activity is positively correlated to the proportion of 1T type with more metallic characteristics[72,73,75-78]. Besides, many workers[77,78] attached importance to the value of 1T/2H interface, indicating that this interface cannot only boost the electron mobility from the electrode to the active sites, but also tune the hydrogen adsorption energy and narrow the band gap for more striking HER behavior.

**2.4 The contributions of edge sites, intrinsic basal sites and crystal phases**

Although the roles of active sites at edges, the intrinsic activity of sites on basal plane and crystal phases have been investigated for the optimal design of electrocatalysts, it is necessary to provide a comprehensive understanding of these factors for enhanced HER activity. Based on the previous discussions above, numerous research focusing on the nanostructure design and optimization has been developed and exploited following three fundamental strategies, namely, increasing the densities of active sites, activating the inert basal plane and 2H-1T phase transformation. In addition, owing to the atom-level thickness and highly enriched active sites of 2D layered $MoX_2$, regulating the electronic band structure to lower the dimensions can evidently lead to greater electron conductivity, better tuned hydrogen adsorption and the optimized local electron density distribution[79]. Hence, identifying the roles of these factors comprehensively could provide a better comprehension of the structure-activity regime of $MoX_2$ electrocatalyst materials.

With an as-prepared mesoporous 1T-type nanosheet-like structure, a systematic study on the roles of active sites, S vacancies (considered as basal sites) and phases of various $MoS_2$ samples, consisting of mesoporous 2H-phase $MoS_2$ (P-2H-$MoS_2$), mesoporous 1T-phase $MoS_2$ (P-1T-$MoS_2$), mesoporous 2H-phase $MoS_2$ after S compensation (P-2H-$MoS_2$+S), 1T-phase $MoS_2$ (1T-$MoS_2$), and 2H-phase $MoS_2$ (2H-$MoS_2$), was performed to identify the priority of the importance of these factors, as **Fig. 12** illustrates[80]. For phase structures, the 1T type (including 1T and P-1T) always exhibits better HER performance than the corresponding 2H type (including 2H and P-2H) because of better metallic properties (e.g. electron conductivity); for the density of active sites, P-1T and P-2H samples show more outstanding activity than the bulk, 1T and 2H counterparts due to more enriched edge sites for accelerated charge transfer kinetics; for the intrinsic in-plane activity (the role of S vacancies as defect sites), S compensation indicates a greater overpotential, in turn showing that the defect active sites of P-2H-MoS$_2$ (vs P-2H-MoS$_2$+S) can enhance the HER performance by decreasing the overpotential. Considering the synergistic enhancement in HER activity, the net effect of edge sites, S vacancies and 1T phases leads to the optimum activity compared with the effect of any single or double factors. However, one drawback in this comprehensive analysis is the lack of the contributions of three factors individually to the intrinsic activity.

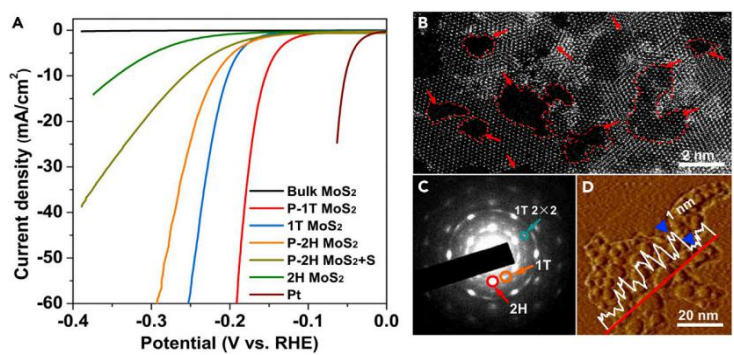

**Fig. 12 (A)** Polarization curves after iR correction indicating a comprehensive comparison in different electrocatalytic performance of various MoS$_2$ samples and Pt (control group); **(B-D)** images of the mesoporous 1T-MoS$_2$ nanosheets using **(B)** high-resolution scanning transmission electron microscopy (HRSTEM), **(C)** selected area electron diffraction pattern, and **(D)** atomic force microscopy (AFM)[80].

Consequently, the role of phase transition is factually closely linked with the increased number of active edge sites and the activated in-plane sites, thus bringing about better conductivity. In other words, the overall effect of crystal phases, edge sites and basal defects depends on the combinations of the density and intrinsic activity of sites (both at edges and on the basal plane).

## 3. Similarities and interconnections of the individual layered molybdenum dichalcogenide: more insights into this family with tunable properties

### 3.1 General analysis of the interconnected properties in MoX$_2$

As discussed in the above sections, the family of molybdenum dichalcogenides are a host of cost-effective, highly efficient, robust, earth-abundant and scalable HER electrocatalysts to replace Pt-based ones. In the form of MoX$_2$, S, Se and Te belong to the congeners named after chalcogens, largely determining the similar physical and structural properties and subsequently reflecting great potentialities in becoming competitive candidates in HER. Besides the roles of crystal phases, edge sites and in-plane activity resembling each other, the electronic properties of chalcogens in excellent catalytic performance is also analogous. With a certain extent of electronegativity, chalcogen atoms can extract electrons from the Mo metals as electron capture sites; also, they could regulate the local electron densities and tailor the band structure of Mo and tune the water dissociation via the $X^{\delta-}$-$Mo^{n+}$-$H_2O$ network through in-plane vacancy engineering[81,82]. As a result, three kinds of $MoX_2$ exhibit relatively great electron mobility, ultrathin thickness, 2D nanosheet-like morphology and large surface area for better catalytic activity and stability.

In addition to the common advantages for potential high-activity catalysts, the underlying rules of the steadily evolved properties of the individuals (disulfides, diselenides and ditellurides) is also of great significance. Here, these versatile but interconnected properties originate from the combinations of Mo and different X elements. Among them, the increased degree of covalency in transition metal-chalcogen (Mo-X) bond and the decreased electronegativity from sulphur to tellurium directly stimulates the enhancement of metallic properties in chalcogen[2,23,83,84]. Also, more distinct metallic properties indicate a lower band gap for reduced energy barriers of water activation and faster kinetics of the Heyrovsky step in HER[84,85]. Thus, the differentiation of elemental features of different chalcogens is another key factor to explore the correlations between $MoX_2$ species and the individual catalytic activity.

**3.2 Examples of $MoTe_2$ electrocatalysts: demonstration of a comparable performance to disulfides and diselenides**

Compared with oxides, sulfides and selenides, $MoTe_2$ materials are theoretically most promising electrocatalysts to replace Pt due to their appealing catalytic activity from the optimal metallic nature, although only several products have been well documented. In this context, this section is to list some typical $MoTe_2$-based HER-related studies to both demonstrate a comparable performance to the other chalcogenides and identify the roles of various forms of defects and active sites.

Algoli et al. [86] completed a comparative study of the HER activity among MoSe$_2$, MoTe$_2$ ultrathin films (with a low coverage of around 1 μg/cm$^2$ to boost the sensitivity to the basal plane) and their tertiary solid solutions (MoSe$_{2-x}$Te$_x$) on the highly oriented pyrolytic graphite (HOPG) substrates. To investigate the role of the emerging grain boundaries on the basal plane of the solid solutions, they made a controllable synthesis of these samples with nearly the same surface morphology in guarantee of no other kind of defects or edge sites. Then from the overall HER test the activity of the as-prepared samples can be comprehensively compared as shown in **Fig. 13**.

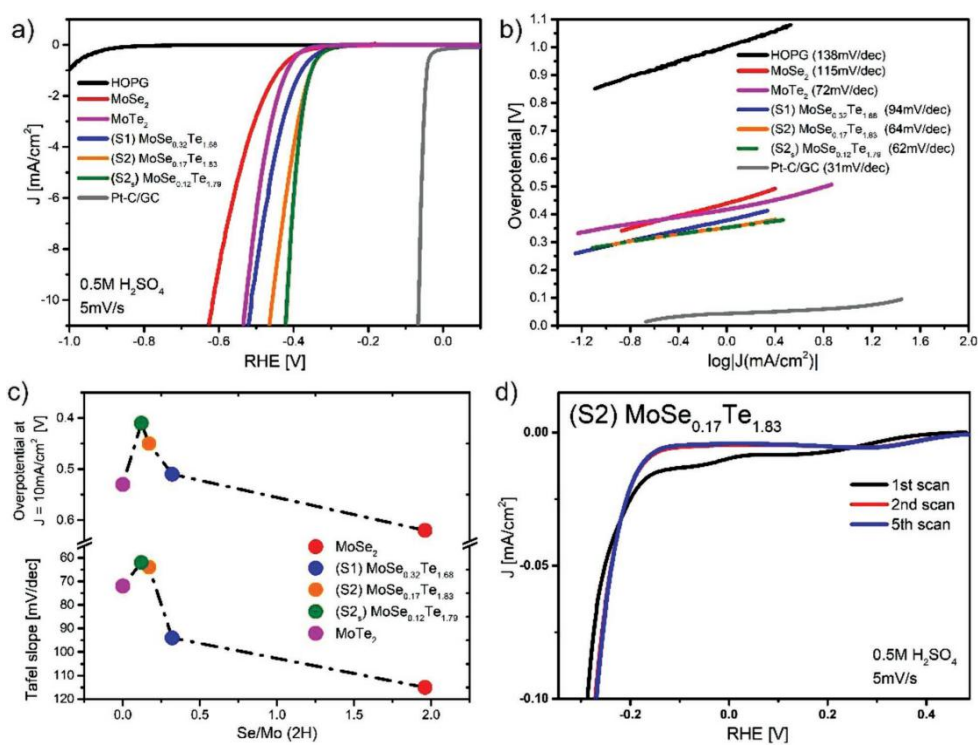

**Fig. 13** HER activity of different molybdenum dichalcogenide electrocatalysts and Pt-C/GC (control group). **(a)** Polarization curves at a scan rate of 5 mV/s in 0.5 M H$_2$SO$_4$. Here, S1 and S2 refers to the metallic solid solutions in different content of Se and Te, and S2$_s$ is obtained from the S2 solution after a mild sputtering treatment; **(b)** corresponding Tafel plots; **(c)** volcano plots showing the overpotential-Se/Mo (2H) ratio dependence at current density of 10 mA/cm$^2$ and the Tafel slope-Se/Mo (2H) ratio dependence, respectively; **(d)** sequence of polarization curves for S2 during 5 scan circles[86].

Considering the solid solutions (S1 and S2) and the pristine MoX$_2$ (X=Se, Te) samples, the most valuable discovery is that the formation of twin boundaries can significantly activate the inherently inert 2H basal plane based on a markedly lower overpotential and Tafel slope of two solid solutions from the volcano-like activity shown in **Fig.13(c)**. More importantly, the contribution of this boundary to HER performance, acting as line defects, can be regulated with the altered Se/Mo ratio. Although the appropriate value of the ratio for the greatest contribution cannot still be obtained, to our delight, the specific formed solid solution with any arbitrary Se/Mo ratio exhibits better electrocatalytic activity than two pure thin-films ($MoSe_2$ and $MoTe_2$).

Furthermore, Cho et al.[87] fabricated a novel hybrid structure with only minimal amount of Pt nanoparticles coated on the atomically monoclinic $MoTe_2$, presenting a superior HER performance with a Tafel slope of 22 mV/dec and an exchange current density of 1.0 $mA/cm^2$ by forming a Pt-Mo alloy layer and enriching more active basal sites.

Until now, although the mechanism comprehension and optimal design of $MoTe_2$-related electrocatalysis are still in the early stage, the similar morphology and better metallic nature of ditellurides can further accelerate the progress of research in this field.

## 4. About hydrothermal: a facile, easily tunable and generalized method

In the section **1**, in spite of the extra water adsorption and dissociation steps competing against the hydrogen adsorption for the goal of the rate determining step in HER, especially when the proton source is not so sufficient, a balanced strategy in both lowering the activation barrier of dissociated water (the kinetics-based aspect) and accelerating the electron mobility on the basal plane (the thermodynamics-based aspect) has been summarized. Then, the section **2** and **3** highlight the creativity and scalability of the nanostructure design of the $MoX_2$ family with respect to the promoted HER rates, activity and stability; specifically, fully considering the similar and interconnected features of disufides, diselenides and ditellurides stemming from their comparable metallic natures and similarly well-modified structural morphologies, a clear mapping on the structure-activity correlations becomes the focus of the oriented, rational design of high-performance catalysts.

For hydrothermal synthesis, it is a wet-chemical techniques to potentially achieve a more comprehensive structure-activity relationship for desired morphology, phases and compositions of $MoX_2$[30,31]. By adjusting several key parameters like temperature, reagents, precursors, solvents and reaction time, it's likely to realize the as-formed products with controlled structural features[31,32,88]. However, the precise control of these expected structures and structure-based properties is a tough and time-consuming task, a comparative study on summarizing how these experimental parameters work for better tuned morphologies at nanoscale and subsequently enhanced electrochemical performance.

Under this circumstance, by extracting and employing three parameters (the species and concentration of precursors, the concentration of reductants and reaction temperature), a one-pot hydrothermal method is proposed based on different experimental systems to comprehensively demonstrate how vital the difference of the surfaces and structures can make for HER enhancement[89-92]. The general process is shown below as well as a schematic diagram presented in **Fig.14**:

The typical synthesis of the $MoSe_2$ product is conducted with the mixture of 0.2 mmol Se metal powder (as-prepared in hydrazine), 0.1 mmol $Na_2MoO_4·2H_2O$ and 0.2 mmol $NaBH_4$. Then, they are dissolved into 80 mL distilled water. The solution is then stirred magnetically for nearly an hour. After that, the solution is transferred into an autoclave, which is then heated in an oven at a temperature-rising speed of 24 ℃/min ultimately to 200 ℃. Then, the autoclave is immersed at the ultimate temperature for 12 h and cooled down to room temperature. Then, the collection of the as-synthesized product is finished via centrifugation at around 5,000 rpm. Finally, after being washed with deionized water and absolute ethanol three times and vacuum-dried at 40 ℃, the product is successfully obtained.

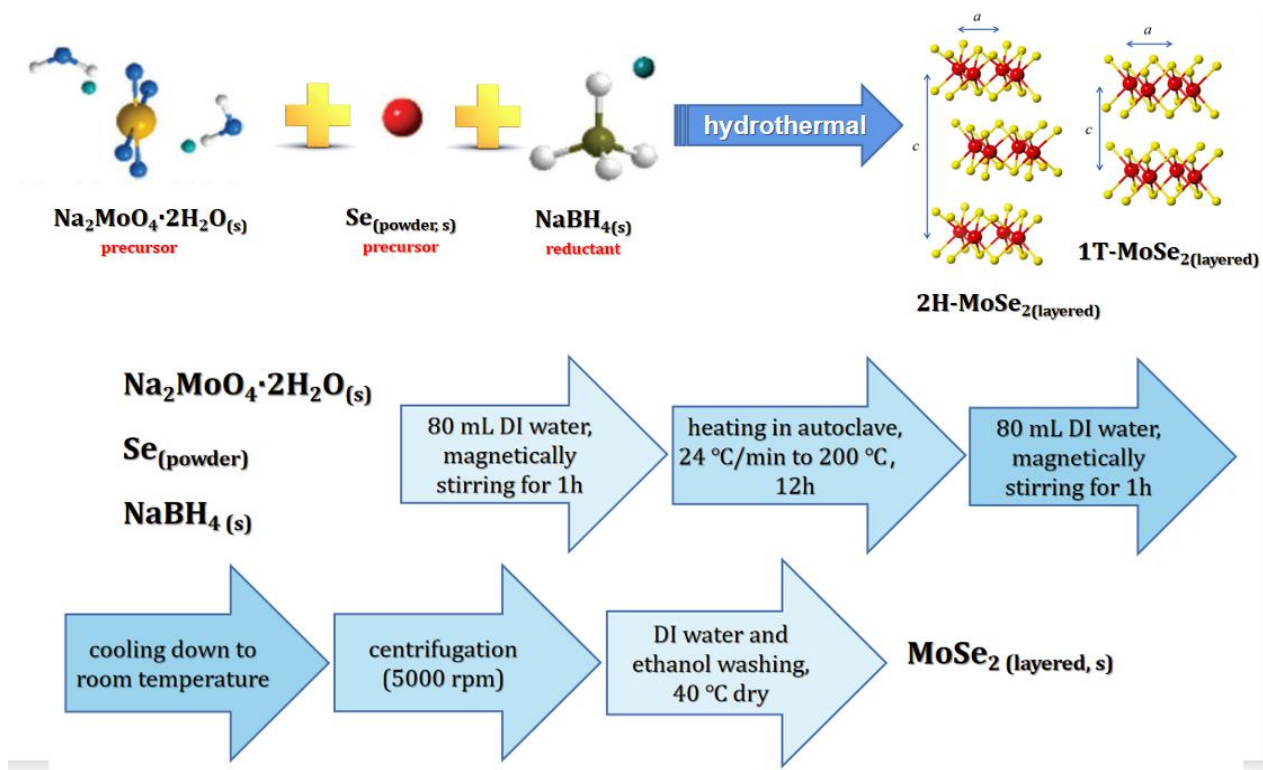

**Fig. 14** schematic illustration of the general hydrothermal synthetic method and the corresponding experiment procedures in detail (listing the synthesis of MoSe$_2$ as an example).

Besides its facile, straight-forward and time-saving procedures, two major advantages of this experimental system are worth mentioning: the flexible tunability and broa generalizability. For its tunability, more explicit structure-activity regimes can be obtained by adjusting the relevant parameters and simultaneously remaining the other irrelevant parameters the same. For example, the synthesis of nano-shaped MoS$_2$ and MoTe$_2$ can be successfully achieved merely by substituting selenium powders with sulphur or tellurium powders as the metal precursor. For its generalization ability, this approach can be applicable under various experimental conditions with a wide-range fluctuation like a series of molybdenum dichalcogenides.

## 5. Evaluation I: the effect of different chalcogenides

This work is to evaluate the effect of different chalcogenides (MoX$_2$ species, including MoS$_2$, MoSe$_2$ and MoTe$_2$) on the HER performance[93]. Explicitly, different members in MoX$_2$ family corresponds to the unique nanosurfaces and nanostructures, thereby bringing about different electrocatalytic performance as catalysts.

## 5.1 Structural characterization

First, powder X-ray diffraction (PXRD) techniques are used for the crystal structure analysis based on different synthesized $MoX_2$ nanostructures. **Fig.15** comparatively depicts three kinds of nanoshaped patterns: the hexagonal $MoS_2$ nanograins, the hexagonal $MoSe_2$ nanoflowers (with sheet-like petals) and the well ordered $MoTe_2$ nanotubes with weaker Te rings, which are all in agreement with the standard crystal structure from database (e.g. JCPDS card no. 00-037-1492 ($MoS_2$), no. 029-0914 ($MoSe_2$) and no. 01-073-1650 ($MoTe_2$)). Notably, the well patterned $MoTe_2$ samples with high order of crystallinity may lower the activation barriers of the formation of the $MoTe_2$ layer attached to the outer surface owing to the weaker Te rings as active nucleation sites.

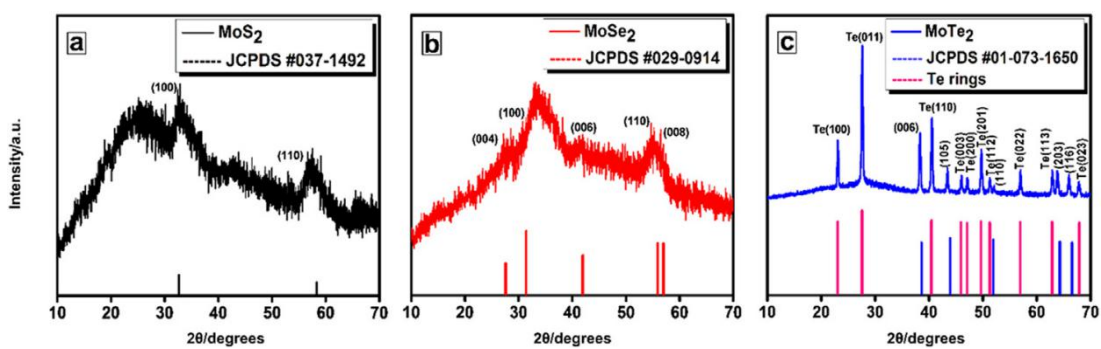

**Fig. 15** PXRD patterns of **(a)** $MoS_2$ nanograins, **(b)** $MoSe_2$ nanoflowers and **(c)** $MoTe_2$ nanotubes[93].

Second, FESEM (field emission scanning electron microscopy) and EDS (Energy-dispersive X-ray spectroscopy) tests are used to provide a nanoscale illustration of structural morphogies and elemental composition of the three categories of molybdenum dichalcogenide nanostructures. The FESEM image in **Fig.16(a)** portrays the poorly-ordered, grain-like $MoS_2$ structures at a 16-20 nm level; conversely, from **(b)** and **(c)**, the sheet-like diselenides and tube-like ditellurides both possess well-ordered shapes. In addition, EDS spectral information depicts the high purity of three as-synthesized nanostructures; the corresponding graphs from elemental mapping also indicate the uniformly-patterned elemental distributions of three structures.

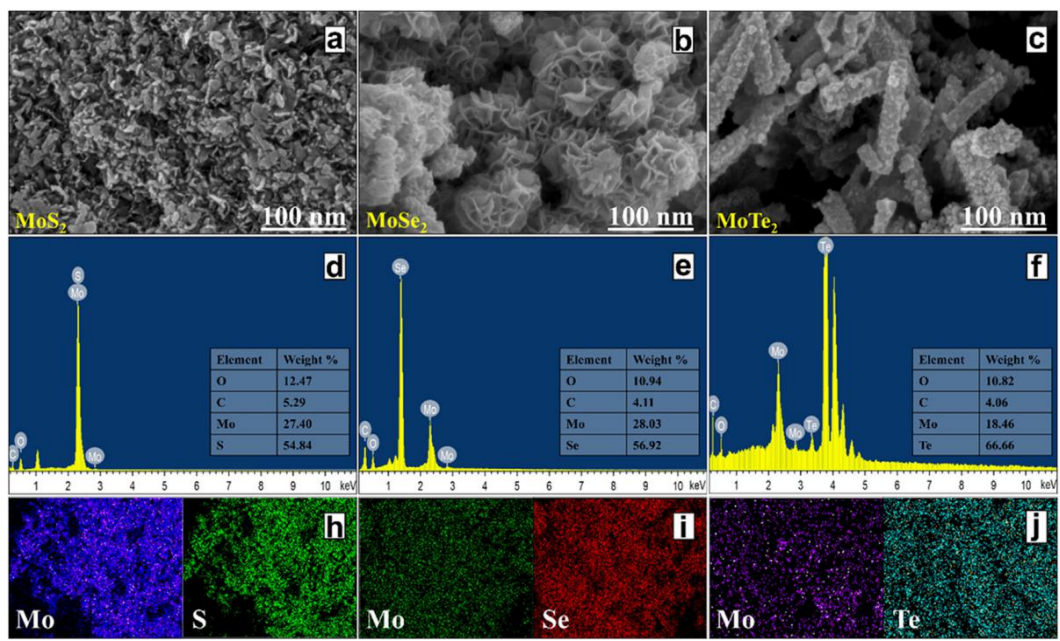

**Fig. 16 (a-c)** PFESEM images of **(a)** MoS$_2$ nanograins, **(b)** MoSe$_2$ nanoflowers and **(c)** MoTe$_2$ nanotubes, showing the individual morphological features; **(d-f)** EDS spectra's illustrations of **(d)** MoS$_2$ nanograins, **(e)** MoSe$_2$ nanoflowers and **(f)** MoTe$_2$ nanotubes, representing their individual elemental purity; **(h-j)** the corresponding elemental mapping results of **(h)** MoS$_2$ nanograins, **(i)** MoSe$_2$ nanoflowers and **(j)** MoTe$_2$ nanotubes, indicative of the uniformity of their elemental distribution[93].

### 5.2 Structure-activity relationship

For the comparison among the HER activity of three nanostructures, the values of overpotential, Tafel slope and double layer capacitance ($C_{dl}$) are tested and further analyzed. MoSe$_2$ is an extraordinarily promising catalyst with the strongest activity among the three, inferred from the optimal overpotential $\eta$ (-330 mV, for 100 mA/cm$^2$ current densities, **Fig.17(a)**) and Tafel slope $b$ value (65.92 mV/dec, **Fig.17(b)**). Furthermore, MoTe$_2$ presents the most accessible electrochemically active surface area (ECSA) for real hydrogen evolution with a tiptop $C_{dl}$ value of 4.57 mF/cm$^2$ (estimated from the equation $2C_{dl} = d(\Delta j)/dV$, where $\Delta j$ stands for the difference in double-layer charging current densities, $V$ means the scan rate, and $d(\Delta j)/dV$ indicates the slope of the corresponding plot showing the $\Delta j$-$V$ dependence in **(c)**[94]). Notably, the lower slope of MoSe$_2$, indicative of its potentially inferior catalytic performance, is partly different from the estimated intrinsic activity from $\eta$ and $b$; the underlying reason is perhaps the nanotube-like structure of MoTe$_2$ provides more widely distributed electrochemically active sites[95]. Additionally, for HER kinetics, the superiority of MoSe$_2$ layered nanosheets as a promising catalysts can be further verified with the lowest charge transfer resistance (showing the best eletrode-electrolyte contact for electron conductivity[96], thereby commensurate with its optimal catalytic performance) in Nyquist plots (**Fig.17(d-f)**).

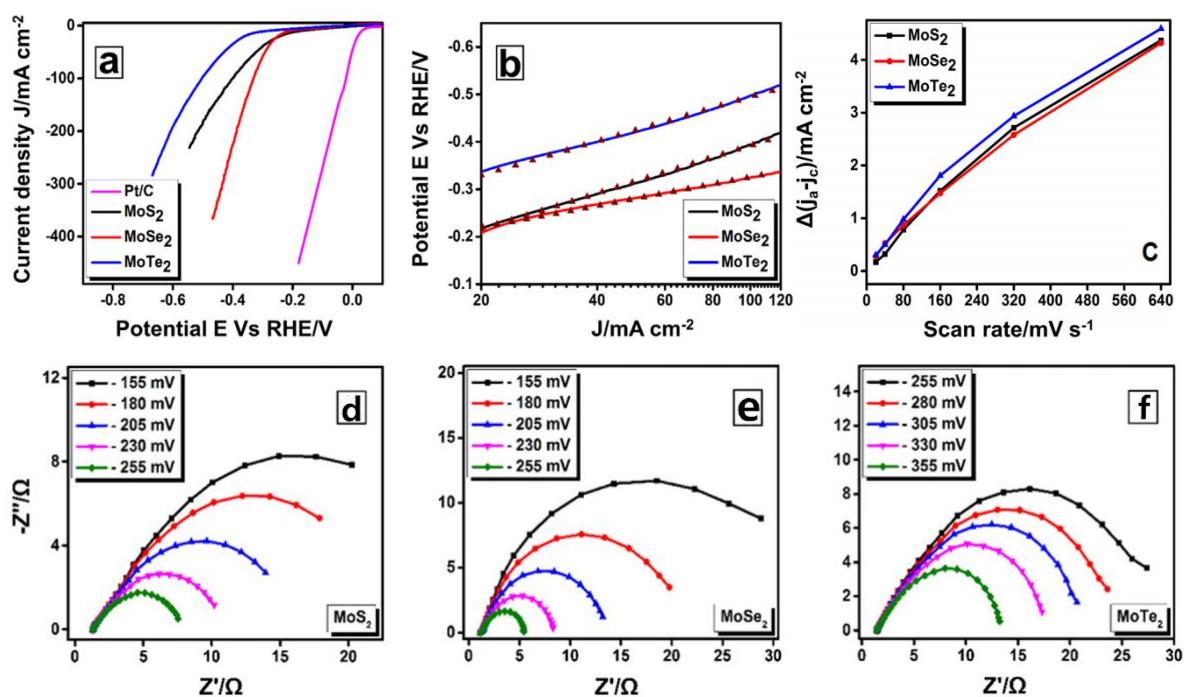

**Fig. 17 (a)** Polarization curves (based on iR corrected LSV) of three MoX$_2$ nanostructures; **(b)** corresponding Tafel plots; **(c)** Plot showing the dependence between the double-layer current densities difference and the scan rate for three nanostructures;; **(d-f)** Nyquist plots of **(d)** MoS$_2$ nanograins, **(e)** MoSe$_2$ nanoflowers and **(f)** MoTe$_2$ nanotubes[93].

For HER stability evaluations via twelve-hour chronopotentiometric (E-t) tests (**Fig. 18**), three categories of nanostructures are demonstrated to be comparably durable electrocatalysts. In detail, the tests of the MoS$_2$ and MoSe$_2$ nanostructures show a slightly escalation maybe due to their sluggish degradation stemming from the S/Se ion bleach; conversely, the long-term stability test of MoTe$_2$ exhibits an upward trend, illustrating the activation of enriched basal sites in the nanotube and nearly no-degradation.

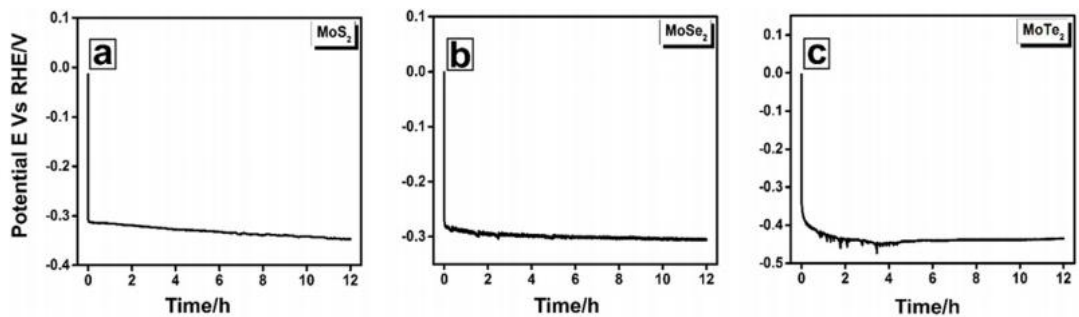

**Fig. 18** E-t curves from the chronopotentiometric stability tests of **(a)** MoS$_2$ nanograins, **(b)** MoSe$_2$ nanoflowers and **(c)** MoTe$_2$ nanotubes[93].

5.3 Summary

Different species of molybdenum dichalcogenides, determining their distinctive nanostructures, all evidently boost HER with the order in performance (catalytic activity): MoSe$_2$ > MoS$_2$ > MoTe$_2$, despite several contributory characteristics for MoTe$_2$ (substantial, spatially distributed active sites, no-degradation). The sheet-like layers of MoSe$_2$ have superior electron conductivity for accelerated HER kinetics owing to its huge surface area and ultrathin thickness.

It can be also inferred that, for a better revelation of structure-activity regimes, we only focus on the as-prepared MoSe$_2$ samples with preferential nanosheet-like structures because of their optimum catalytic activity, satisfactory robustness and well-modified traits.

## 6. Evaluation II: the effect of different concentration of metal precursors

This work is to evaluate the effect of different metal precursor concentration on the HER performance[26,97,98]. The parameter can be regulated with different stoichiometric chalcogen/Mo ratios (Se source: Se powder and Mo source: Na$_2$MoO$_4$), so five samples (expressed in the form of MoSe$_x$) with the Se/Mo ratios $x$ = 1.8, 2.0, 2.2, 2.3 and 2.4 were prepared.

6.1 Structure-activity relationship

In terms of the electrocatalytic activity of 5 samples from **Fig. 19**, MoSe$_{2.3}$ displays the lowest overpotential of 0.130 V (at the current densities of 10 mA/cm$^2$) and Tafel slope of 46 mV/dec, which is commensurate with its sensational intrinsic catalytic activity. Also, the lowest charge-transfer resistance (11.61 Ω, from **Table 1**) for MoSe$_{2.3}$ depicts its rapidest electron mobility for HER kinetics. With regard to the HER stability, the chronoamperometric (CA) test of MoSe$_{2.3}$ even suggests better robustness with a decrease of $\Delta j$ = 1 mA/cm$^2$ than that of commercial Pt/C declined by 2 mA/cm$^2$ (**Fig.19(c)**). Likewise, the LSV results in **(d)** before and after 12h of CA test verifies the excellent long-term stability of MoSe$_{2.3}$.

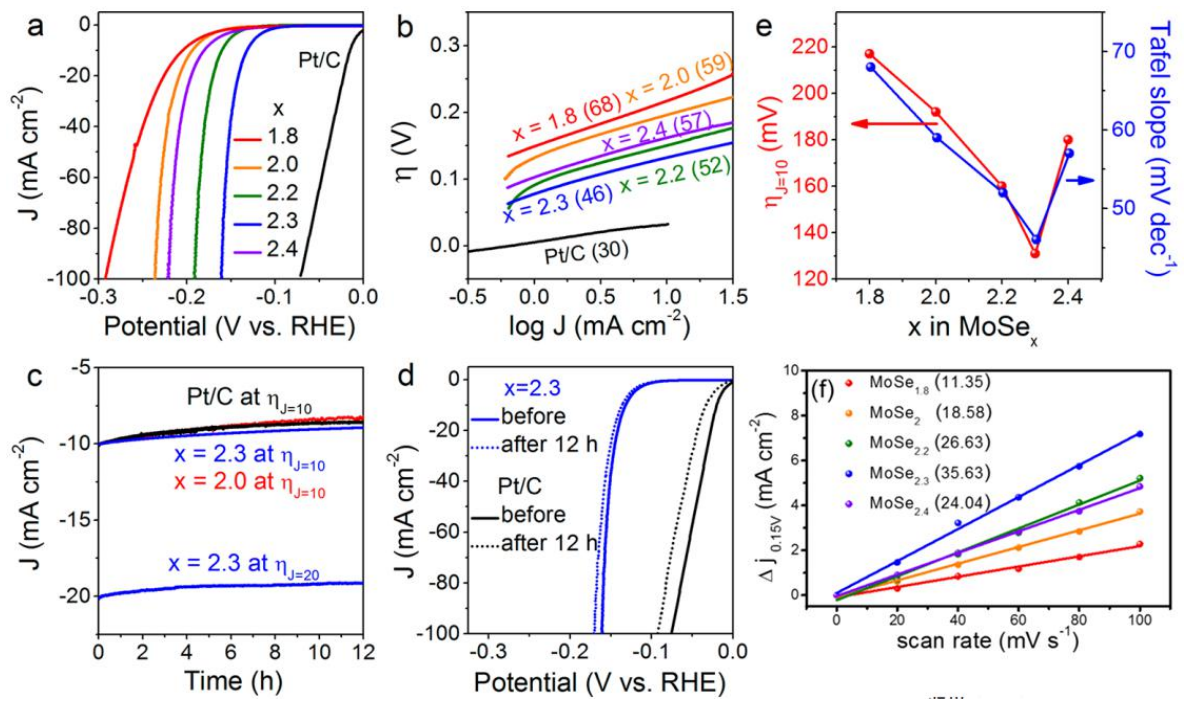

**Fig. 19 (a)** Polarization curves of five MoSe$_x$ samples (x = 1.8, 2.0, 2.2, 2.3, 2.4); **(b)** corresponding Tafel plots with the Tafel slope of each sample shown in parentheses; **(c, d)** HER stability of MoSe$_x$ nanosheets: **(c)** twelve-hour chronoamperometric (CA) curves of Pt/C at η$_{J=10}$, MoSe$_2$ at η$_{J=10}$, and MoSe$_{2.3}$ at η$_{J=10}$ and η$_{J=20}$; **(d)** polarization curves of MoSe$_{2.3}$ and Pt/C before and after 12 h of CA test; **(e)** plots of the η$_{J=10}$-x and Tafel slope-x dependence; **(f)** Plot showing the dependence between the difference in double-layer current densities and the scan rate, with the $R_{CT}$ (charge-transfer resistance) of each sample displayed in parentheses, where the slope ($\Delta j/\Delta V$) indicates the $2C_{dl}$ value[97,98].

Table 1. Characteristics of five MoSe$_x$ samples and their parameters for HER performance

|   | Se/Na$_2$MoO$_4$ ratio | [Se]/[Mo] ratio | $\eta_{j=10}$/(V vs RHE) | $b$/(mV·dec$^{-1}$) | $R_{CT}$/Ω | $C_{dl}$/(mF·cm$^{-2}$) |
|---|---|---|---|---|---|---|
| 1 | 1.5 | 1.8 | 0.217 | 68 | 24.54 | 11.35 |
| 2 | 1.75 | 2.0 | 0.192 | 59 | 20.92 | 18.58 |
| 3 | 2.0 | 2.2 | 0.160 | 52 | 13.14 | 26.63 |
| 4 | 2.25 | 2.3 | 0.130 | 46 | 11.61 | 35.63 |
| 5 | 2.5 | 2.4 | 0.180 | 57 | 16.24 | 24.04 |

More importantly, with the [Se]/[Mo] ratio (*x*) increases from 1.8 to 2.3, enhancement in HER activity are presented in both **Table 1** and **Fig.19(e)**. This is because of the [Se]/[Mo] ratio-induced 2H-to-1T' phase conversion, where a 2H phase-dominated structure exists at *x* = 1.8 or 2.0 and when *x* exceeds 2.2, a formed superlattice structure with the heterogeneity of the 2H type and a new 1T' phase type emerges, signified in XRD patterns (**Fig.20(a)**). In addition, the emergence of the 1T' phase is demonstrated to occur at *x* = 2.2 accompanied by the weakening of 1H phase signal and the formation of Se-Se bond (**Fig.20(b)**). Furthermore, for more insight into the contribution of 1T' phase, XPS data (**Fig.20(c-e)**) is obtained to investigate the function between the fraction of 1T' phase and the HER activity. In detail, as *x* increases, the fraction of 1T'-type rises: 13, 25, 64, 71, and 75% for *x* = 1.8, 2.0, 2.2, 2.3, and 2.4, respectively.

Based on the systematic structural-activity analysis around **Fig.20**, the promoted HER activity with the increased *x* is ascribed to the higher fraction of 1T' phase with more metallic properties. Compared with the semiconducting 2H phase, 1T' phase renders adequate active sites, metallic conductivity and better electrode kinetics[33,46,57]. Note that the inferior performance of MoSe$_{1.8}$ with Se vacancies to that of MoSe$_2$ may disobey the previous discoveries that Se vacancies can facilitate the HER performance as active sites[99,100], so the main contribution for HER enhancement may still come from the percentage of the metallic 1T'-type in the phase structure. Also notably, the declined electrocatalytical performance of MoSe₂.₄ is probably because of the precipitation of excess Se, which are separated from the layered nanosheets, restricting the electron mobility and hindering the HER kinetics, as presented in **Fig.21**.

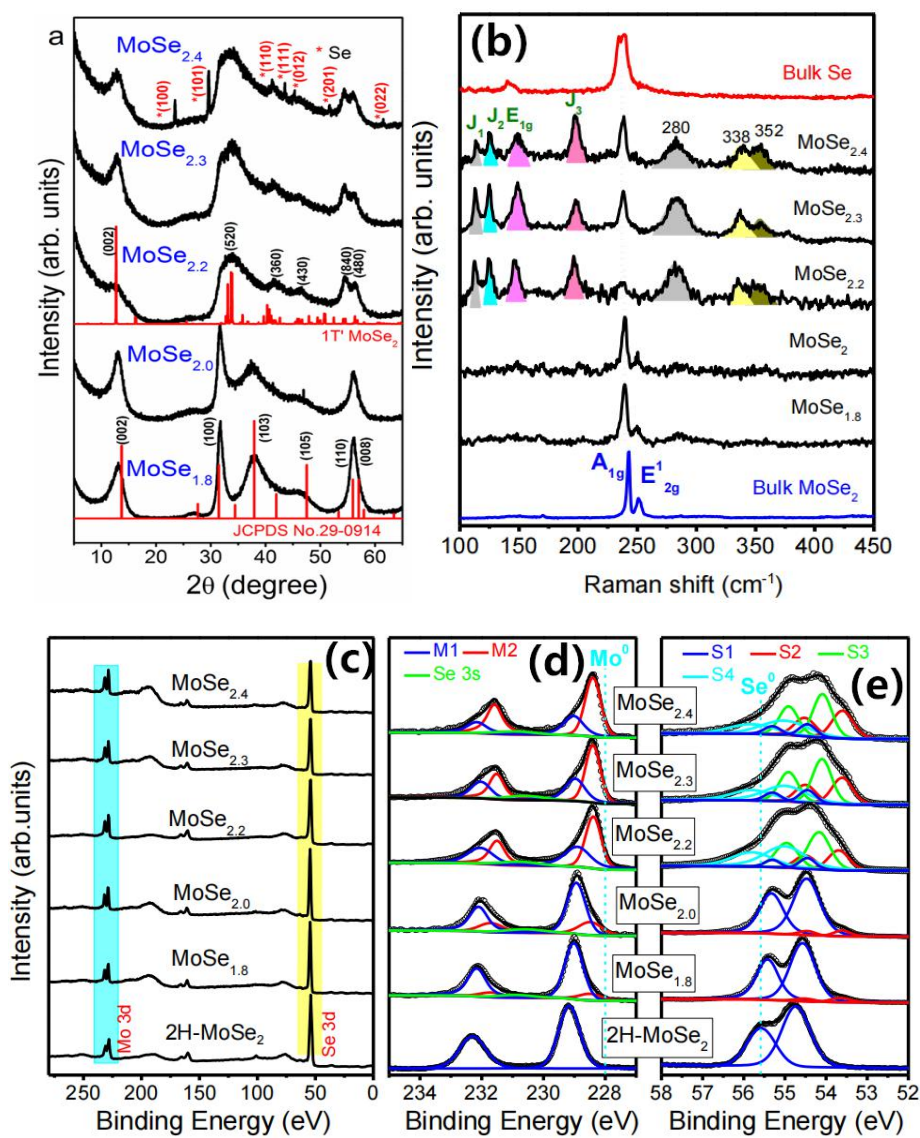

**Fig. 20 (a)** XRD patterns of MoSe$_x$ with x = 1.8, 2.0, 2.2, 2.3, and 2.4, showing their crystal structures. The peaks are referenced to 2H and 1T' phase MoSe$_2$, where at x = 2.4 the signal of separated Se appears; **(b)** Raman spectrum of MoSe$_x$ samples, bulk 2H-MoSe$_2$ and bulk Se. The excitation source is 532 nm diode laser. To describe, the two characteristic signal peaks of bulk 2H-MoSe$_2$, corresponding to the in-plane A$_{1g}$ and out-of-plane E$^1_{2g}$ vibration modes, are shown at x = 1.8 and 2. The other MoSe$_x$ samples (x = 2.2, 2.3, 2.4) show the unique Raman peaks of 1T' phase: J$_1$, J$_2$, J$_3$ and E$_{1g}$ mode. Also, as *x* increases, the stronger peaks always appear where the bulk Se peak also appears (238 nm$^{-1}$), so this peak comes from the Se-Se bond. The peaks at 280, 338, and 352 cm$^{-1}$ could originate from the Se-Se modes; **(c)** XPS survey scans, **(d)** fine-scan Mo 3d, and **(e)** fine-scan Se 3d peaks of MoSe$_x$ and bulk 2H-MoSe$_2$. For more comprehension, **(c)** the area ratio of Se/Mo peak increases with x; **(d)** the 2H-MoSe$_2$ shows the 2H phase 3$d_{5/2}$ peak at 229.2 eV, blue-shifted from the Mo metal at 228.0 eV. The peak of MoSe$_x$ was resolved into the 1T' phase band at 228.4 eV and the 2H phase at 228.9 eV. The 2H-type is the main phase at x = 1.8 and 2.0, and the 1T'-type coexisting with 2H-type serves as the main phase at x = 2.2, 2.3 and 2.4; **(e)** The Se 3$d_{5/2}$ peak of 2H-MoSe$_2$ at 54.7 eV, 1 eV red-shifted from the neutral Se (Se$^0$) signal at 55.6 eV, corresponds to the 2H-phase Se$^{2-}$ anions bonded with the Mo cations. For x = 1.8 and 2.0, the major peaks (S1 band) at 54.5 eV match the 2H-phase Se$^{2-}$ anions while the minor ones (S2 band) at 53.6 eV, 2 eV red-shifted from the Se$^0$, are assigned to the 1T'-phase Se$^{2-}$ anions. For x = 2.2, 2.3 and 2.4, the wide-range peak are resolved into 4 bands: the 2H-phase Se$^{2-}$ (S1 band), the 1T'-phase Se$^{2-}$ (S2 band), the 1T'-phase Se$_2^{2-}$ bridge anions (S3 band), and the 2H-phase Se$_2^{2-}$ anions (S4 band)[98].

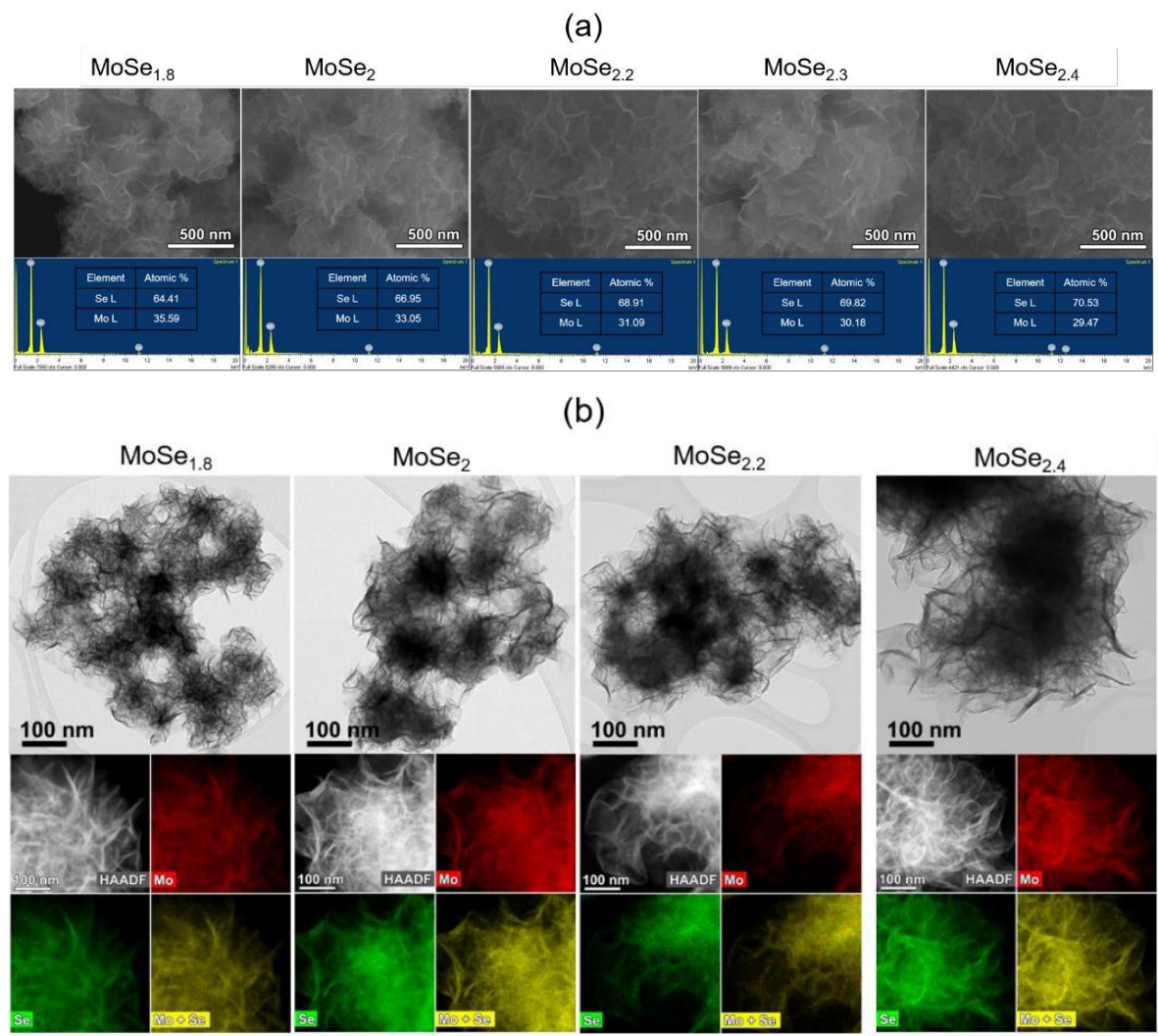

**Fig. 21 (a)** SEM and EDX; **(b)** HRTEM images, HAADF-STEM images and elemental mapping of MoSe$_x$. They all have sheet-like nanoflower-structured spheres, but the flower-like sphere contains the Se phases separately at x = 2.4 [97,99].

For better predictions in the structure of Se-rich 1T'-MoSe$_2$ with the potential HER pathway, it's of great significance to ascertain what kind of new intralayer and/or interlayer Se-bonds emerge in hydrothermal-guided HER optimization. The contributions of excess Se are concluded with the identification of the optimized energy-lowest Se-rich model[98,101]: (1) substituting the Mo atoms; (2) forming the interstitial bonds adjacent to the substituted Se atoms; (3) forming the interlayer bridge adjacent to the substituted Se. Also, the interstitial Se adatoms have been pinpointed to be the most electroactive to HER because of the strikingly lessened activation barrier of H* adsorption[98].

**6.2 Summary**

The 2H-to -1T' phase transition and control of MoSe$_x$ can be realized with the regulation of [Se]/[Mo] ratio $x$ during the hydrothermal synthesis, thereby bringing about the prompted HER activity: as $x$ increases, the fraction of 1T' phase increases. If no segregated Se phase exists, the excess Se (normally x > 2) can bring the Se-rich MoSe$_2$ more metallic properties by optimizing the electron conductivity in bulk counterparts. MoSe$_{2.3}$ demonstrates the optimal electrocatalytic performance with the lowest overpotential of 0.130 V and the Tafel slope of 46 mV/cm$^2$. Structurally, the interstitially bonded Se adatoms, used as the substitution of Mo, can act predominantly for HER enhancement.

**7. Evaluation III: the effect of different concentration of reductants**

This work is to evaluate the effect of different metal precursor concentration on the HER performance[102,103]. The parameter can be regulated with the different stoichiometric [B]/[Mo] (B source: NaBH$_4$ and Mo source: Na$_2$MoO$_4$) ratios $y$, so four samples (expressed in the form of MoSe$_2$-$y$) with different [B]/[Mo] ratios $y$ = 1, 2, 3 and 4 were prepared.

**7.1 Structure-activity relationship**

With regard to the electronic structure of the MoSe$_2$-$y$ samples, Raman and XPS data

(**Fig.22**) both display an explicit 2H-to-1T phase transition process with altered electronic structure and induced atomic rearrangement, which is similar to the process stimulated by different ratios of B metal precursor. The same, the underlying regulation mechanism between the concentration of NaBH$_4$ reductant and the crystal phases is: as *y* increases, the fraction of 1T phase increases. In detail, among the samples from *y* = 1 to 4, the area (intensity) of 2H-phase characteristic peaks is weakened while the area of 1T-phase peaks gets stronger, indicating the increasing y value benefits the structural conversion from 2H- to 1T-phase MoSe$_2$[104].

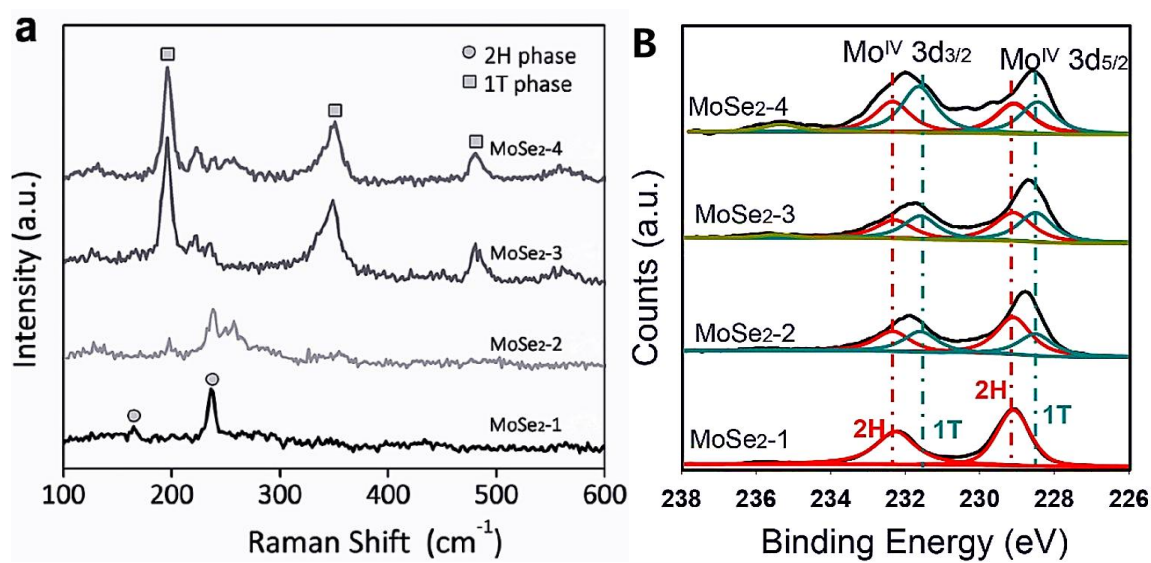

**Fig. 22 (a)** Raman spectra of MoSe$_2$-*y* nanosheets synthesized by adding different concentration of NaBH4 reductant; **(b)** Mo 3d XPS spectra of MoSe$_2$-*y* nanosheets[102].

In addition to more bulk conductivity for metallic 1T-phase type, they have one more morphological advantages compared with the 2H type illustrated in TEM and HRTEM images (**Fig. 23**). The TEM image depicts a few-layer nanosheet structure with ultrathin thickness and broad surface area, suggesting a great potential to be high-activity electrocatalysts; further, the expanded interlayer spacing (0.82 nm, exceeding that of 2H-MoSe$_2$ at 0.7 nm) of (002) plane for MoSe$_2$ samples can be clearly observed under higher resolution, accordingly increasing the exposure of active defect sites.

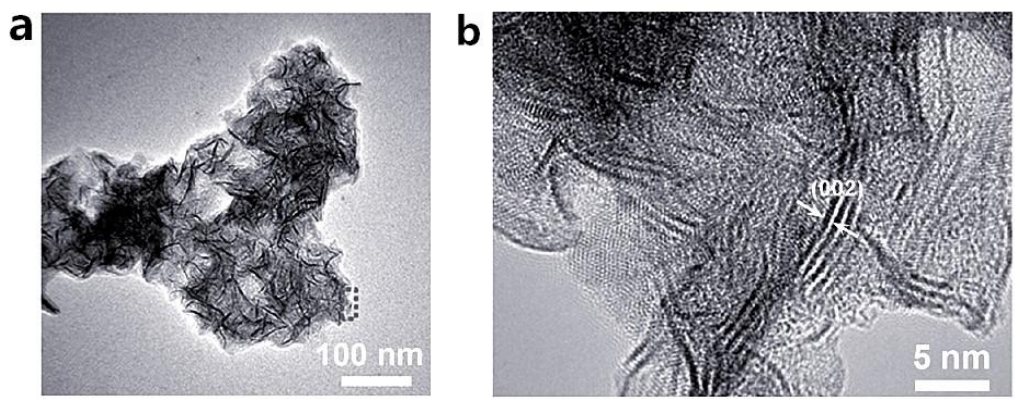

**Fig. 23 (a)** TEM, and **(b)** HRTEM images of MoSe$_2$-4 samples, indicating **(a)** a 3-5 monolayers and **(b)** a highly expanded (002) plane of 1T-MoSe$_2$[102].

Predictably, the MoSe$_2$-4 sample reveals the superior HER performance with an appealing overpotential of 163 mV (vs RHE) under the current density of 10 mA/cm$^2$ and a low Tafel slope of 55 mV/dec, firmly demonstrating the positive effect of a higher content of 1T phase on the electrocatalytic performance.

**7.2 Summary**

The higher [B]/[Mo] ratio contributes to the 2H-to-1T phase transition by causing a higher content of metallic 1T phase, which strikingly facilitates the hydrogen reaction. The existence of 1T phase exposes more unsaturated defects as active sites by enlarging the interlayer spacing and also accelerates the bulk conductivity by in-plane activation. The optimum HER activity is obtained based on the appropriately high amount of the added NaBH$_4$.

## 8. Evaluation IV: the effect of different reaction temperatures

This work is to evaluate the effect of different metal precursor concentration on the HER performance[90,102]. The parameter can be regulated with the different terminal temperature manually set in autoclave (e.g. 200 °C as the target terminal temperature shown in the hydrothermal procedure of **Section 4**).

**8.1 Structure-activity relationship: rough predictions of the optimal hydrothermal temperature**

As the first step, to predict roughly the appropriate reaction temperature in autoclave during the hydrothermal synthesis, a series of parallel experiments was set at the temperature of 200, 300, 400 °C[90]. Compared with the formation of the high-performance, nanosheet-like MoSe$_2$ with the expanded interlayer spacing (0.82 nm), the 200 °C-synthesized lamellar MoSe$_2$ possess the comparable interlayer expansion of 1.16 nm, an indication of the potential to be more electroactive with more enriched active sites, shown in the HRTEM image of **Fig. 24**. Specifically, the sheet-like layers are all assembled into the flower-like spheres at edges and internal basal plane area; the highly interlayer-expanded nanosheets are composed of several monolayers. These are in good agreement with the previous research [15,25,49,51,53,97,102,105] and the relevant images in **Section 7**. However, when the reaction temperature is over 200 °C, the HRTEM image reveals the typical 2H bulk state with the interlayer spacing at 0.58 nm. It can be inferred that the temperature at around 200 °C during the hydrothermal process can lead to a 2H-to-1T phase transition accompanied by the interlayer expansion, besides the excess ratio of reactants (e.g. precursors and reducing agencies). What's more, as the temperature reaches 300 °C or higher, the measured interlayer spacing shows the value appearing in the typical 2H-MoSe$_2$, suggesting a possible 1T-to-2H restoration at higher hydrothermal temperature.

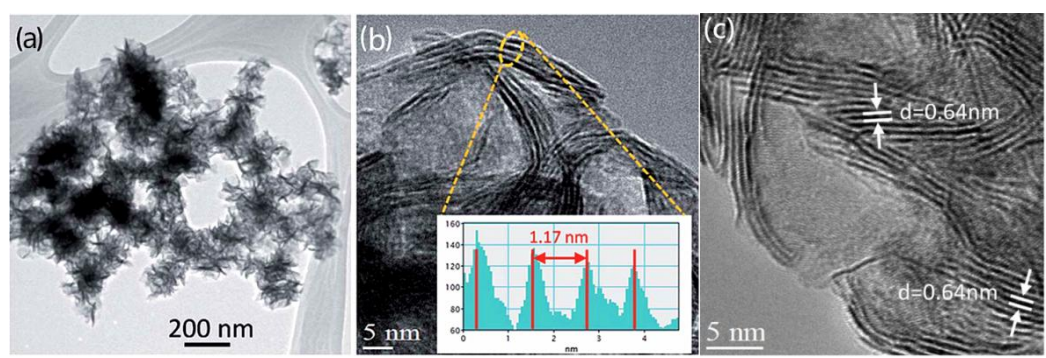

**Fig. 24 (a)** TEM image of 200 °C-synthesized MoSe$_2$ samples, showing a flower-like structure with 3-7 monolayers; **(b)** HRTEM image of 200 °C-synthesized MoSe$_2$ samples with a highly expanded interlayer spacing of 1.17 nm, shown in the inset calibrated profile plot; **(c)** HRTEM image of 300 °C-synthesized MoSe$_2$ samples, showing a few-layer sheet-like structure with an interlayer spacing of 0.64 nm [90].

To verify the above hypotheses, more structural and morphological analysis are further required. XRD patterns (**Fig.25(a)**) of the as-formed samples at different hydrothermal temperatures provide a comprehensive comparison among the characteristic peaks of 1T- and 2H-phase. For MoSe$_2$ samples at 200 °C and the calculated pattern of expanded MoSe$_2$, a stronger peak at 2θ = 7.6° and a weaker peak at 2θ = 15.2° represent the *d*-spacing of 1.17 nm and 0.58 nm, respectively, further demonstrated by the correspondence between the doubling peak height at 2θ = 7.6° and the doubling interlayer spacing value (1.17 nm ≈ 2×0.58 nm)[106]. Likewise, the (002) peak at 2θ = 13.7° for samples synthesized at 300 or 400 °C is attributed to the typical spacing at 0.64 nm in bulk 2H-MoSe$_2$; the other peaks are all indexed to the 2H-phase samples (JCPDF 29-0914). Consequently, a hydrothermal temperature at approximately 200 °C can endow MoSe$_2$ with the metallic 1T phase type and a expanded structure while the MoSe$_2$ nanostructures can be restored to the 2H bulk counterparts at 300 °C and higher.

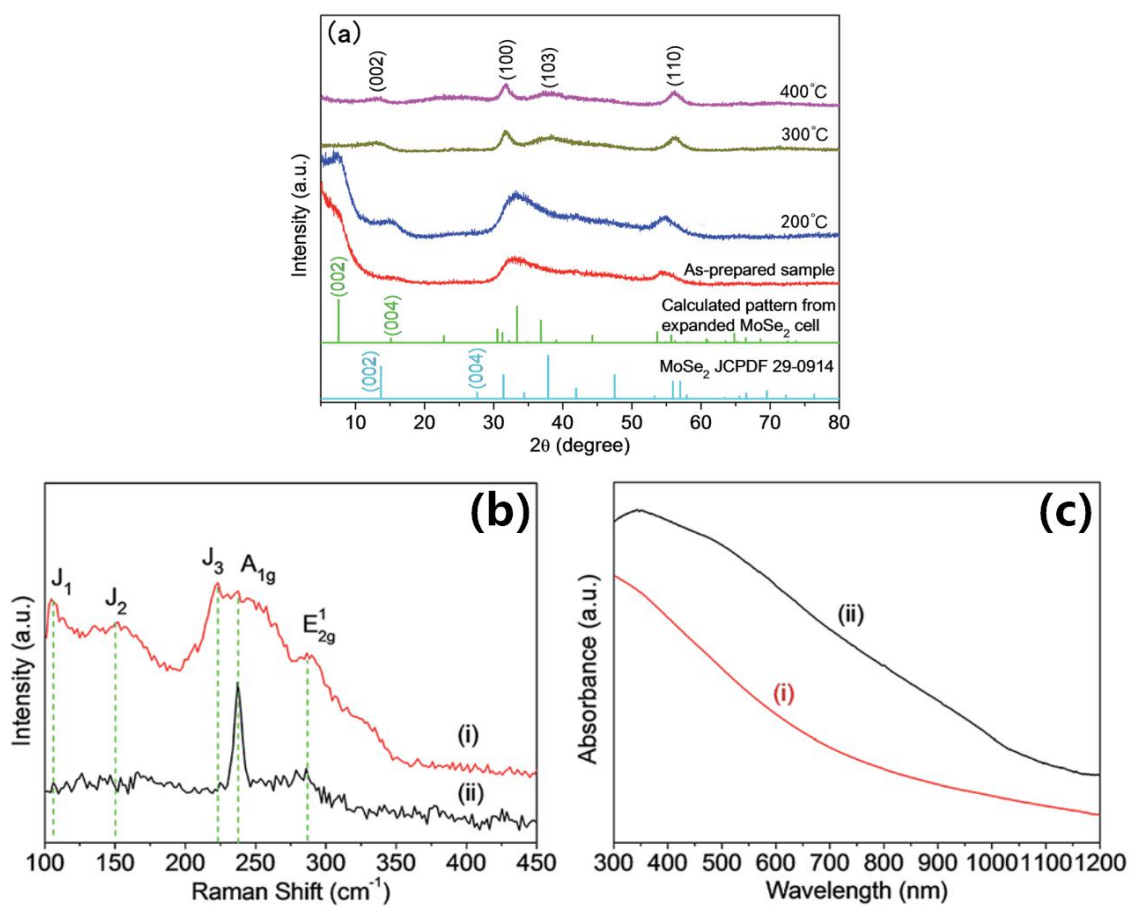

**Fig. 25 (a)** XRD patterns of MoSe$_2$ samples as-prepared, calculated and synthesized at various hydrothermal temperatures, showing a marked 1T-to-2H phase conversion when the temperature is over 200 °C; **(b)** Ramen spectra and **(c)** UV-vis absorption spectra of (i) as-prepared 1T- and (ii) 300 °C-synthesized 2H-MoSe$_2$ samples[90].

The phase transformation information can also be verified based on the Raman and UV-vis absorption spectra in **Fig.25(b-c)**. There are three additional clear XRD signals for the as-prepared 1T-type, identified as the three kinds of phonon modes: $J_1$ (106 cm$^{-1}$), $J_2$ (149 cm$^{-1}$) and $J_3$ (221 cm$^{-1}$) modes. However, only two peaks at 237 cm$^{-1}$ and 286 cm$^{-1}$ for 2H-type can be indexed to the in-plane $A_{1g}$ and out-of-plane $E_{2g}^1$ mode, respectively; these two resonance peaks also appear in (i), demonstrating the co-existence of 1T and 2H phase in the 200 °C-synthesized sample (in detail, 1T-type: the major phase and 2H-type: the minor one). These patterns well verifies the distinct structural features between 1H- and 2T-phase as well as different interlayer distances. In addition, two kinds of crystal phases suggest distinct light-absorption properties, shown in UV-vis images: a marked red shift from the peak at 700nm for the as-prepared 1T-phase sample to that at 1000 nm for the synthesized 2H-phase sample. The shift means a direct band gap of 1.4-1.7 eV based on the monolayer nanostructure, also related to the expanded interlayer distance[107].

After the hypotheses from observation and the validation based on structural analysis, the role of different phases due to the rising hydrothermal temperature are directly evaluated electrochemically, between the 200 °C- and 300 °C-synthesized samples. From **Fig.26**, referring to the considerably low Tafel slope and tiny charge-transfer resistance, the 1T-MoSe$_2$ synthesized at 200 °C delivers a higher HER performance, suggesting the appropriate temperature is around 200 °C.

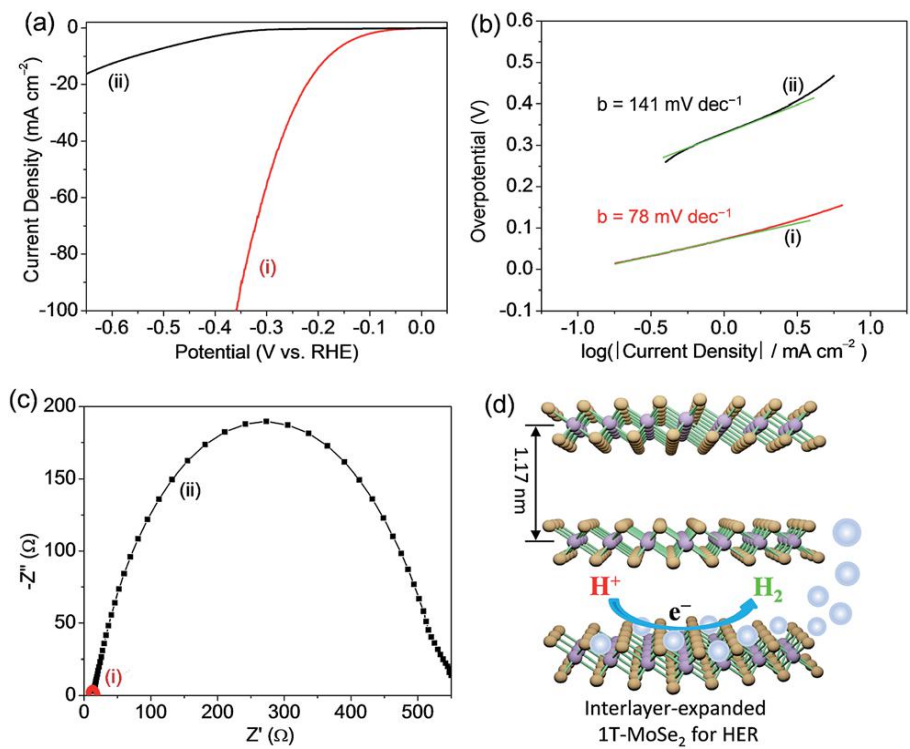

**Fig. 26 (a)** Polarization curves, **(b)** correponding Tafel plots and **(c)** Nyquist plots of (i) 200 °C-synthesized 1T-MoSe$_2$ and (ii) 300 °C-synthesized 2H-MoSe$_2$ samples; **(d)** schematic illustration of the HER behavior in the interlayer-expanded 1T-MoSe$_2$ nanosheets[90].

**8.2 Structure-activity relationship: accurate estimations of the optimal hydrothermal temperature**

The above rough estimation identifies the influence of 2H-to-1T conversion at 200 °C and the conversion back to 2H-phase at higher hydrothermal temperatures; further, more accurate estimations of the appropriate temperature attach importance to the role of the altered fraction of 1T-phase in tuning the HER activity. In this context, smaller temperature difference of 20 °C was set at the range of 140~200 °C, and accordingly, the MoSe$_2$ samples synthesized at different temperatures are expressed as MoSe$_2$-*T*.

Similarly to the detailed analysis in **Fig.20**, Raman spectroscopy is used to ascertain the phase transition with spectral illustration of different content of each phase type and XPS is used to further quantify the content. As can be inferred from **Fig.27 (a)**, a noticeable 2H-to-1T phase transition occurs with the increased hydrothermal temperature, where the 2H phase plays the predominant role in MoSe$_2$-140 and MoSe$_2$-160 samples while the only peaks associated with the MoSe$_2$-180 and MoSe$_2$-200 samples corresponds to the 1T-type. In **Fig.27(b)**, the intensity of 1T- and 2H-MoSe$_2$ peaks can be derived from the deconvolution of Mo 3$d$ peaks. The individual contents of samples generated at various temperatures are summarized in the **Table 2.** below. The increased temperature leads to the larger fraction of 1T-phase from 140 °C to 180 °C, and in-turn the decreased 1T-phase to 200 °C, which implies an excessively high temperature may induce partially a conversion from 1T- back to 2H-phase[29,35,90]. As a result, MoSe$_2$-180 is confirmed to possess the largest content of metallic 1T phase.

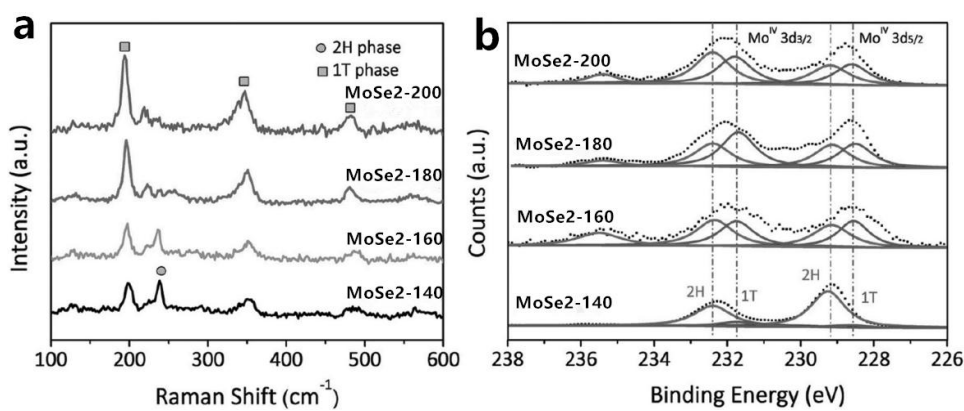

**Fig. 27 (a)** Raman, **(b)** and Mo 3d EDX spectra of MoSe2-*T* produced at various temperatures[102].

**Table 2.** Summary of the phase content

| MoSe$_2$-*T* | Percentage of Crystal Phase types | |
|---|---|---|
| | 2H | 1T |
| 140 | 91 | 9 |
| 160 | 57 | 43 |
| 180 | 45 | 55 |
| 200 | 52 | 48 |

Then the practical HER performance of these MoSe$_2$ samples at different temperatures via hydrothermal treatment was measured with a fixed loading area of 0.14 ± 0.01 mg·cm$^{-2}$. After iR correction, the MoSe$_2$-180 sample, possessing the highest fraction of 1T phase, reaches the lowest overpotential of 152 mV vs RHE at $j$ = -10 mA/cm$^2$ and the exceptional Tafel slope of 52 mV/dec, perfectly indicating the well-performed intrinsic HER activity. To further understand the charge transfer kinetics, the double-layer capacitance was measured to precisely obtain the real availability of electrocative surface area. Likewise, MoSe$_2$-180 exhibits the minimal $C_{dl}$ (27.8 mF/cm$^2$) and the maximal $R_{ct}$ (16 Ω). This gives a powerful explanation that the available active sites contain not only the edge sites in inert 2H bulk state, but also the in-plane basal sites, strongly signifying the predominant role of 1T crystal phase in facilitating the HER reaction rate and activity. Additionally, the HER parameters of the other samples are portrayed in the **Table 3** below.

Table 3. Summary of a system of HER parameters

| MoSe$_2$-$T$ | $\eta_{j=10}$/(mV vs RHE) | $b$/(mV·dec$^{-1}$) | $C_{dl}$/(mF·cm$^{-2}$) | $R_{ct}$/(Ω) |
|---|---|---|---|---|
| 140 | 211 | 72 | 1.25 | 58 |
| 160 | 197 | 54 | 14.7 | 36 |
| 180 | 152 | 52 | 27.8 | 16 |
| 200 | 163 | 55 | 25.2 | 23 |

For more structure-activity comprehension, there is an surprising discrepancy between the Raman spectra and XRD-based calculations (**Fig.27**), the frontier revealing the information on the pristine 1T phase of MoSe$_2$-180 and MoSe$_2$-200 while the mixture of 1T and 2H phases elucidated from the latter. The missing information about the 2H phase may be ascribed to the possible core/shell-like structure (1T phase as the shell and 2H phase as the core). In detail, the shorter sampling depth of the Raman detection cannot insert the internal 2H phase. To further verify this assumption, reaction time is naturally reckoned as another factor linked with the structures/interfaces of MoSe$_2$ nanosheets. Consequently, an extra time-dependent study of the effect on HER activity needs to be carried out, where the core idea is how the increased reaction time contributes to modulating the phase structure. The corresponding recent discoveries [102,108,109] showed the longer time induces the conversion to 1T phase (specifically, the pure 2H phase for only several hours, the coexistence of mixed phases for around 8~10 hours, and the pure 1T phases for over 12 hours).

### 8.3 Summary

The determination and validation of the optimal hydrothermal temperature for HER enhancement are mainly divided into two steps: the first is for rough research aiming at the approximate value of temperature (around 200 °C); the second is to further ascertain the accurate value of temperature based on the first step (180 °C).

The appropriate hydrothermal temperature is verified to be 180 °C. The MoSe$_2$-180 sample delivers the most appealing HER activity and performance with the lowest overpotential at a current density of 10 mA/cm$^2$, the lowest Tafel slope and the minimal charge-transfer resistance. The underlying mechanism is that the optimal temperature leads to the greatest content of metallic 1T phase. Further, considering the role of 1T phase for HER enhancement, the phase is suggested to play a predominant role compared with the defects (disorders) and bulk conductivity.

## 9. Conclusion and perspectives

The low concentration of proton donors in alkaline HER, subsequently leading to the extra water adsorption and dissociation steps, identifies the value of active sites (edge and basal sites) and crystal phases in lowering the extra activation barrier and/or optimizing the H* adsorption kinetics; in addition, the outstanding morphology-based features (surface area, thickness, defects, disorders and crystallinity) of layered molybdenum dichalcogenide families pinpoint the roles of active sites and phases for more interpretable and feasible structure-activity analysis. In this context, hydrothermal synthetic method is used to exhibit a clear mapping between the nanostructure/nanosurface design and the practical HER performance by adjusting key experimental parameters. In this article, MoX$_2$ nanostructures in different species (X = S, Se, Te), the molar ratio of added reactants (the Se metal precursor and the NaBH$_4$ reducing agency) and hydrothermal temperature are considered for the modulated structure and the optimized HER performance.

The tunability of the hydrothermal method can be well confirmed with regard to its structure-activity relationship and the underlying mechanism. A system of MoX$_2$-based samples delivery their excellent HER activity, stability and kinetics, with the optimal value of overpotential $\eta$, exchange current density $j_0$, Tafel slope $b$ and charge transfer resistance $R_{ct}$, which are well tuned by these parameters above. For better comprehensions of the parameter-tunable structure-activity correlations, the role of active sites and phases are crucially highlighted. In detail, different chalcogenide species are indicative of different exposure of surface defects as active sites on nanoscale; the concentration of the added precursor/reductant determines the specific content of metallic 1T/1T' phase by inducing a 2H-to-1T(1T') conversion; hydrothermal temperatures regulates the phase and defect structure simultaneously by generating a controlled core/shell-like structure with mixed phases.

Furthermore, the tunable procedures contribute to more revelation in the weigh of the roles of structural factors (edge sites, bulk conductivity for in-plane activation and phases). The crystal phase plays the predominant role as the phase transition also results in the altered densities of active sites and intrinsic activity of basal plane.

To conclude, with higher tunability and scalability, the hydrothermal method can pave a novel path for the oriented, rational design of higher-activity transition-metal-based electrocatalysts and better understandings of the underlying design rules and mechanisms.

# References

[1]  Zheng, Y.; Jiao, Y.; Jaroniec, M.; Qiao, S. Advancing the Electrochemistry of the Hydrogen Evolution Reaction through Combining Experiment and Theory. *Angew. Chem. Int. Ed.* **2015**, *54*, 52-65.
[2]  Liu, Y.; Wu, J.; Hackenberg, K. P.; Zhang, J.; Wang, Y. M.; Yang, Y.; Keyshar, K.; Gu, J.; Ogitsu, T.; Vajtai, R. Self-Optimizing, Highly Surface-Active Layered Metal Dichalcogenide Catalysts for Hydrogen Evolution. *Nat. Energy* **2017**, *2*, 17127.
[3]  Jaramillo, T. F.; Jørgensen, K. P.; Bonde, J.; Nielsen, J. H.; Horch, S.; Chorkendorff, I. Identification of Active Edge Sites for Electrochemical H$_2$ Evolution from MoS$_2$ Nanocatalysts. *Science* **2007**, *317*, 100-102.
[4]  Gillis, R. J.; Al-Ali, K.; Green, W. H. Thermochemical Production of Hydrogen from Hydrogen Sulfide with Iodine Thermochemical Cycles. *Int. J. Hydrog. Energy* **2018**, *43*, 12939-12947.
[5]  Zou, X.; Zhang, Y. Noble Metal-Free Hydrogen Evolution Catalysts for Water Splitting. *Chem. Soc. Rev.* **2015**, *44*, 5148-5180.
[6]  Tarasevich, M. R.; Korchagin, O. V. Electrocatalysis and pH (a Review). *Russ. J. Electrochem.* **2013**, *49*, 600-618.

[7] Auinger, M.; Katsounaros, I.; Meier, J. C.; Klemm S. O.; Biedermann, P. U.; Topalov, A. A.; Rohwerdera, M.; Mayrhofer, K. J. Near-Surface Ion Distribution and Buffer Effects During Electrochemical Reactions. *Phys. Chem. Chem. Phys.* **2011**, *13*, 16384-16394.

[8] Strmcnik, D.; Uchimura, M.; Wang, C.; Subbaraman, R.; Danilovic, N.; Van Der Vliet, D.; Paulikas, A. P.; Stamenkovic, V. R.; Markovic, N. M. Improving the Hydrogen Oxidation Reaction Rate by Promotion of Hydroxyl Adsorption. *Nature Chem.* **2013**, *5*, 300-306.

[9] Zhu, S.; Qin, X.; Yao, Y.; Shao, M. pH-Dependent Hydrogen and Water Binding Energies on Platinum Surfaces as Directly Probed through Surface-Enhanced Infrared Absorption Spectroscopy. *J. Am. Chem. Soc.* **2020**, *142*, 8748-8754.

[10] Danilovic, N.; Subbaraman, R.; Strmcnik, D.; Chang, K. -C.; Paulikas, A. P.; Stamenkovic, V. R.; Markovic, N. M. Enhancing the Alkaline Hydrogen Evolution Reaction Activity through the Bifunctionality of $Ni(OH)_2$/Metal Catalysts. *Angew. Chem. Int. Ed.* **2012**, *51*, 12495-12498.

[11] Cheng, T.; Wang, L.; Merinov, B. V.; Goddard III, W. A. Explanation of Dramatic pH-Dependence of Hydrogen Binding on Noble Metal Electrode: Greatly Weakened Water Adsorption at High pH. *J. Am. Chem. Soc.* **2018**, *140*, 7787-7790.

[12] Yang, X.; Nash, J.; Oliveira, N.; Yan, Y.; Xu, B. Understanding the pH Dependence of Underpotential Deposited Hydrogen on Platinum. *Angew. Chem. Int. Ed.* **2019**, *58*, 17718-17723.

[13] Wang, R.; Yan, J.; Zu, M.; Yang, S.; Cai, X.; Gao, Q.; Fang, Y.; Zhang, S. Facile Synthesis of Interlocking g-$C_3N_4$/CdS Photoanode for Stable Photoelectrochemical Hydrogen Production. *Electrochim. Acta* **2018**, *279*, 74-83.

[14] Shi, X.; Fields, M.; Park, J.; McEnaney, J. M.; Yan, H.; Zhang, Y.; Tsai, C.; Jaramillo, T. F.; Sinclair, R.; Norskov, J. K.; Zheng, X. Rapid Flame Doping of Co to $WS_2$ for Efficient Hydrogen Evolution. *Energy Environ. Sci.* **2018**, *11*, 2270-2277.

[15] Wang, L.; Tsang, C.; Liu, W.; Zhang, X.; Zhang, K.; Ha, E.; Kwok, W. M.; Park, J. H.; Lee, L. Y. S.; Wong, K. Y. Disordered Layers on $WO_3$ Nanoparticles Enable Photochemical Generation of Hydrogen from Water. *J. Mater. Chem. A* **2019**, *7*, 221-227.

[16] Kanda, Y.; Kawanishi, K.; Tsujino, T.; Al-otaibi, A.; Uemichi, Y. Catalytic Activities of Noble Metal Phosphides for Hydrogenation and Hydrodesulfurization Reactions. *Catalysts* **2018**, *8*, 160.

[17] Miao, J.; Lang, Z.; Zhang, X.; Kong, W.; Peng, O.; Yang, Y.; Wang, S.; Cheng, J.; He, T.; Amini, A.; Wu, Q.; Zheng, Z.; Tang, Z.; Cheng, C. Polyoxometalate-Derived Hexagonal Molybdenum Nitrides (MXenes) Supported by Boron, Nitrogen Codoped Carbon Nanotubes for Efficient Electrochemical Hydrogen Evolution from Seawater. *Adv. Funct. Mater.* **2019**, *29*, 1970046.

[18] Kou, Z.; Wang, T.; Wu, H.; Zheng, L.; Mu, S.; Pan, Z.; Lyu, Z.; Zang, W.; Pennycook, S. J.; Wang, J. Twinned Tungsten Carbonitride Nanocrystals Boost Hydrogen Evolution Activity and Stability. *Small* **2019**, *15*, 1900248.

[19] Wang, X.; Zheng, Y.; Sheng, W.; Xu, Z. J.; Jaroniec, M.; Qiao, S. Z. Strategies for Design of Electrocatalysts for Hydrogen Evolution under Alkaline Conditions. *Mater. Today* **2020**, *36*, 125-138.

[20] Jiao, Y.; Zheng, Y.; Davey, K.; Qiao, S. Z. Activity Origin and Catalyst Design Principles for Electrocatalytic Hydrogen Evolution on Heteroatom-Doped Graphene. *Nat. Energy* **2016**, *1*, 16130.

[21] Skúlason, E.; Jónsson, H. Atomic Scale Simulations of Heterogeneous Electrocatalysis: Recent Advances. *Adv. Phys.* **2017**, *2*, 481-495.

[22] Parsons, R. The Rate of Electrolytic Hydrogen Evolution and the Heat of Adsorption of Hydrogen. *Trans. Faraday Soc.* **1958**, *54*, 1053-1063.

[23] Wang, R.; Han, J.; Zhang, X.; Song, B. Synergistic Modulation in $MX_2$ (where M = Mo or W or V, and X = S or Se) for an Enhanced Hydrogen Evolution Reaction. *J. Mater. Chem. A* **2018**, *6*, 21847-21858.

[24] Deng, S.; Yang, F.; Zhang, Q.; Zhong, Y.; Zeng, Y.; Lin, S.; Wang, X.; Lu, X.; Wang, C.-Z.; Gu, L.; Xia, X.; Tu, J. Phase Modulation of (1T-2H)-MoSe$_2$/TiC-C Shell/Core Arrays via Nitrogen Doping for Highly Efficient Hydrogen Evolution Reaction. *Adv. Mater.* **2018**, *30*, 1802223.

[25] Vikraman, D.; Akbar, K.; Hussain, S.; Yoo, G.; Jang, J.-Y.; Chun, S.-H.; Jung, J.; Park, H. J. Direct Synthesis of Thickness-Tunable MoS$_2$ Quantum Dot Thin Layers: Optical, Structural and Electrical Properties and Their Application to Hydrogen Evolution. *Nano Energy* **2017**, *35*, 101-114.

[26] Guo, W.; Chen, Y.; Wang, L.; Xu, J.; Zeng, D.; Peng, D. -L. Colloidal Synthesis of MoSe$_2$ Nanonetworks and Nanoflowers with Efficient Electrocatalytic Hydrogen-Evolution Activity. *Electrochim. Acta* **2017**, *231*, 69-76.

[27] Wei, J.; Zhou, M.; Long, A.; Xue, Y.; Liao, H.; Wei, C.; Xu, Z. J. Heterostructured Electrocatalysts for Hydrogen Evolution Reaction Under Alkaline Conditions. *Nano-Micro Lett.* **2018**, *10*, 75.

[28] Zheng, Y.; Jiao, Y.; Zhu, Y.; Li, L. H.; Han, Y.; Chen, Y.; Jaroniec, M.; Qiao, S. Z. High Electrocatalytic Hydrogen Evolution Activity of an Anomalous Ruthenium Catalyst. *J. Am. Chem. Soc.* **2016**, *138*, 16174-16181.

[29] Zhang, B.; Liu, J.; Wang, J.; Ruan, Y.; Ji, X.; Xu, K.; Chen, C.; Wan, H.; Miao, L.; Jiang, J. Interface Engineering: the Ni(OH)$_2$/MoS$_2$ Heterostructure for Highly Efficient Alkaline Hydrogen Evolution. *Nano Energy* **2017**, *37*, 74-80.

[30] Zhen, C.; Zhang, B.; Zhou, Y.; Du, Y.; Xu, P. Hydrothermal Synthesis of Ternary MoS$_{2x}$Se$_{2(1-x)}$ Nanosheets for Electrocatalytic Hydrogen Evolution. *Inorg. Chem. Front.* **2018**, *5*, 1386-1390.

[31] Gao, M.; Xu, Y.; Jiang, J.; Yu, S. Nanostructured Metal Chalcogenides: Synthesis, Modification, and Applications in Energy Conversion and Storage Devices. *Chem. Soc. Rev.* **2013**, *42*, 2986-3017.

[32] Liu, Z.; Gao, Z.; Liu, Y.; Xia, M.; Wang, R.; Li, N. Heterogeneous Nanostructure Based on 1T-Phase MoS$_2$ for Enhanced Electrocatalytic Hydrogen Evolution. *ACS Appl. Mater. Interfaces* **2017**, *9*, 25291-25297.

[33] Fu, Q.; Han, J.; Wang, X.; Xu, P.; Yao, T.; Zhong, J.; Zhong, W.; Liu, S.; Gao, T.; Zhang, Z.; Xu, L.; Song, B. 2D Transition Metal Dichalcogenides: Design, Modulation, and Challenges in Electrocatalysis. *Adv. Mater.* **2021**, *33*, 1907818.

[34] Strmcnik, D.; Lopes, P. P.; Genorio, B.; Stamenkovic, V. R.; Markovic, N. M. Design Principles for Hydrogen Evolution Reaction Catalyst Materials. *Nano Energy* **2016**, *29*, 29-36.

[35] Durst, J.; Siebel, A.; Simon, C.; Hasche, F.; Herranz, J.; Gasteiger, H. A. New Insights into the Electrochemical Hydrogen Oxidation and Evolution Reaction Mechanism. *Energy Environ. Sci.* **2014**, *7*, 2255-2260.

[36] Wei, C.; Sun, Y.; Scherer, G. G.; Fisher, A. C.; Sherburne, M.; Ager, J. W.; Xu, Z. J. Surface Composition Dependent Ligand Effect in Tuning the Activity of Nickel-Copper Bimetallic Electrocatalysts toward Hydrogen Evolution in Alkaline. *J. Am. Chem. Soc.* **2020**, *142*, 7765-7775.

[37] Wang, X.; Xu, C.; Jaroniec, M.; Zheng, Y.; Qiao, S. -Z. Anomalous Hydrogen Evolution Behavior in High-pH Environment Induced by Locally Generated Hydroniumions. *Nat. Commun.* **2019**, *10*, 4876.

[38] Cheng, T.; Wang, L.; Merinov, B. V.; Goddard, W. A. Explanation of Dramatic pH-Dependence of Hydrogen Binding on Noble Metal Electrode: Greatly Weakened Water Adsorption at High pH. *J. Am. Chem. Soc.* **2018**, *140*, 7787-7790.

[39] Ruqia, B.; Choi, S. I. Pt and Pt-Ni(OH)$_2$ Electrodes for the Hydrogen Evolution Reaction in Alkaline Electrolytes and Their Nanoscaled Electrocatalysts. *ChemSusChem* **2018**, *11*, 2643-2653.

[40] Feng, X. J.; Wu, J. Q.; Tong, Y. X.; Li, G. R. Efficient Hydrogen Evolution on Cu Nanodots-Decorated Ni$_3$S$_2$ Nanotubes by Optimizing Atomic Hydrogen Adsorption and Desorption. *J. Am. Chem. Soc.* **2018**, *140*, 610-617.

[41] Voiry, D.; Yang, J.; Chhowalla, M. Recent Strategies for Improving the Catalytic Activity of 2D TMD Nanosheets toward the Hydrogen Evolution Reaction. *Adv. Mater.* **2016**, *28*, 6197-6206.

[42] Gushchin, A. L.; Laricheva, Y. A.; Sokolov, M. N.; Llusar, R. Tri-and Tetranuclear Molybdenum and Tungsten Chalcogenide Clusters: on the Way to New Materials and Catalysts. *Russ. Chem. Rev.* **2018**, *87*, 670-706.

[43] Zhong, W.; Xiao, B.; Lin, Z.; Wang, Z.; Huang, L.; Shen, S.; Zhang, Q.; Gu, L. RhSe$_2$: a Superior 3D Electrocatalyst with Multiple Active Facets for Hydrogen Evolution Reaction in Both Acid and Alkaline Solutions. *Adv. Mater.* **2021**, *33*, 2007894.

[44] McGlynn, J. C.; Cascallana-Matías, I.; Fraser, J. P.; Roger, I.; McAllister, J.; Miras, H. N.; Symes, M. D.; Ganin, A. Y. Molybdenum Ditelluride Rendered into an Efficient and Stable Electrocatalyst for the Hydrogen Evolution Reaction by Polymorphic Control. *Energy Technol.* **2018**, *6*, 345-350.

[45] Bhat, K. S.; Nagaraja, H. S. Performance Evaluation of Molybdenum Dichalcogenide (MoX$_2$; X = S, Se, Te) Nanostructures for Hydrogen Evolution Reaction. *Int. J. Hydrog. Energy* **2019**, *44*, 17878-17886.

[46] Wang, H.; Yuan, H.; Hong, S. S.; Li, Y.; Cui, Y. Physical and Chemical Tuning of Two-Dimensional Transition Metal Dichalcogenides. *Chem. Soc. Rev.* **2015**, *44*, 2664-2680.

[47] Hinnemann, B.; Moses, P. G.; Bonde, J.; Jørgensen, K. P.; Nielsen, J. H.; Horch, S.; Chorkendorff, I.; Nørskov, J. K. Biomimetic Hydrogen Evolution: MoS$_2$ Nanoparticles as Catalyst for Hydrogen Evolution. *J. Am. Chem. Soc.* **2005**, *127*, 5308-5309.

[48] Luo, Z.; Ouyang, Y.; Zhang, H.; Xiao, M.; Ge, J.; Jiang, Z.; Wang, J.; Tang, D.; Cao, X.; Liu, C.; Xing, W. Chemically Activating MoS$_2$ via Spontaneous Atomic Palladium Interfacial Doping towards Efficient Hydrogen Evolution. *Nat. Commun.* **2018**, *9*, 2120.

[49] (a) Yu, Y.; Huang, S. -Y.; Li, Y.; Steinmann, S. N.; Yang, W.; Cao, L. Layer-Dependent Electrocatalysis of MoS$_2$ for Hydrogen Evolution. *Nano Lett.* **2014**, *14*, 553-558.; (b) Janik, J. M.; McCrum, I. T.; Koper, M. T. M. On the Presence of Surface Bound Hydroxyl Species on Polycrystalline Pt Electrodes in the "Hydrogen Potential Region" (0-0.4 V-RHE). *J. Catal.* **2018**, *367*, 332-337; (c) Qian, Z.; Jiao, L.; Xie, L. Phase Engineering of Two-Dimensional Transition Metal Dichalcogenides. *Chin. J. Chem.* **2020**, *38*, 753-760.

[50] Khossossi, N.; Singh, D.; Ainane, A.; Ahuja, R. Recent Progress of Defect Chemistry on 2D Materials for Advanced Battery Anodes. *Chem. Asian J.* **2020**, *15*, 3390-3404.

[51] Coleman, J. N.; Lotya, M.; O'Neill, A.; Bergin, S. D.; King, P. J.; Khan, U.; Young, K.; Gaucher, A.; De, S.; Smith, R. J.; Shvets, I. V.; Arora, S. K.; Stanton, G.; Kim, H. -Y.; Lee, K.; Kim, G. T.; Duesberg, G. S.; Hallam, T.; Boland, J. J.; Wang, J. J.; Donegan, J. F.; Grunlan, J. C.; Moriarty, G.; Shmeliov, A.; Nicholls, R. J.; Perkins, J. M.; Grieveson, E. M.; Theuwissen, K.; McComb, D. W.; Nellist, P. D.; Nicolosi, V. Two-Dimensional Nanosheets Produced by Liquid Exfoliation of Layered Materials. *Science* **2011**, *331*, 568.

[52] Kong, D.; Wang, H.; Cha, J. J.; Pasta, M.; Koski, K. J.; Yao, J.; Cui, Y. Synthesis of MoS$_2$ and MoSe$_2$ Films with Vertically Aligned Layers. *Nano Lett.* **2013**, *13*, 1341-1347.

[53] Lukowski, M. A.; Daniel, A. S.; Meng, F.; Forticaux, A.; Li, L. S.; Jin, S. Enhanced Hydrogen Evolution Catalysis from Chemically Exfoliated Metallic MoS$_2$ Nanosheets. *J. Am. Chem. Soc.* **2013**, *135*, 10274-10277.

[54] Vattikuti, S. P.; Devarayapalli, K. C.; Nagajyothi, P. C.; Shim, J. Microwave Synthesized Dry Leaf-Like Mesoporous MoSe$_2$ Nanostructure as an Efficient Catalyst for Enhanced Hydrogen Evolution and Supercapacitor Applications. *Microchem. J.* **2020**, *153*, 104446.

[55] Lei, Z.; Xu, S.; Wu, P. Ultra-Thin and Porous MoSe$_2$ Nanosheets: Facile Preparation and Enhanced Electrocatalytic Activity towards the Hydrogen Evolution Reaction. *Phys. Chem. Chem. Phys.* **2016**, *18*, 70-74.

[56] Jaramillo, T. F.; Jorgensen, K. P.; Bonde, J.; Nielsen, J. H.; Horch, S.; Chorkendorff, I. Identification of Active Edge Sites for Electrochemical H$_2$ Evolution from MoS$_2$ Nanocatalysts. *Science* **2007**, *317*, 100-102.

[57] Nguyen, T. P.; Choi, S.; Jeon, J. M.; Kwon, K. C.; Jang, H. W.; Kim, S. Y. Transition Metal Disulfide Nanosheets Synthesized by Facile Sonication Method for the Hydrogen Evolution Reaction. *J. Phys. Chem. C* **2016**, *120*, 3929-3935.

[58] Ambrosi, A.; Sofer, Z.; Pumera, M. Lithium Intercalation Compound Dramatically Influences the Electrochemical Properties of Exfoliated $MoS_2$. *Small* **2015**, *11*, 605-612.

[59] Attanayake, N. H.; Thenuwara, A. C.; Patra, A.; Aulin, Y. V.; Tran, T. M.; Chakraborty, H.; Borguet, E.; Klein, M. L.; Perdew, J. P.; Strongin, D. R. Effect of Intercalated Metals on the Electrocatalytic Activity of 1T-$MoS_2$ for the Hydrogen Evolution Reaction. *ACS Energy Lett.* **2017**, *3*, 7-13.

[60] Lin, S.; Kuo, J. Activating and Tuning Basal Planes of $MoO_2$, $MoS_2$, and $MoSe_2$ for Hydrogen Evolution Reaction. *Phys. Chem. Chem. Phys.* **2015**, *17*, 29305-29310.

[61] Ouyang, Y.; Ling, C.; Chen, Q.; Wang, Z.; Shi, L.; Wang, J. Activating Inert Basal Planes of $MoS_2$ for Hydrogen Evolution Reaction through the Formation of Different Intrinsic Defects. *Chem. Mater.* **2016**, *28*, 4390-4396.

[62] Vasu, K.; Meiron, O. E.; Enyashin, A. N.; Bar-Ziv, R.; Bar-Sadan, M. Effect of Ru Doping on the Properties of $MoSe_2$ Nanoflowers. *J. Phys. Chem. C* **2019**, *123*, 1987-1994.

[63] Voiry, D.; Mohiteb, A.; Chhowalla, M. Phase Engineering of Transition Metal Dichalcogenides. *Chem. Soc. Rev.* **2015**, *44*, 2702-2712.

[64] Wang, Q. H.; Kalantar-Zadeh, K.; Kis, A.; Coleman, J. N.; Strano, M. S. Electronics and Optoelectronics of Two-Dimensional Transition Metal Dichalcogenides. *Nat. Nanotechnol.* **2012**, *7*, 699-712.

[65] Lukowski, M. A.; Daniel, A. S.; Meng, F.; Forticaux, A.; Li, L.; Jin, S. Enhanced Hydrogen Evolution Catalysis from Chemically Exfoliated Metallic $MoS_2$ Nanosheets. *J. Am. Chem. Soc.* **2013**, *135*, 10274-10277.

[66] Wang, D.; Zhang, X.; Bao, S.; Zhang, Z.; Fei, H.; Wu, Z. Phase-Engineering of Multiphasic 1T/2H $MoS_2$ Catalyst for Highly Efficient Hydrogen Evolution. *J. Mater. Chem. A* **2017**, *5*, 2681-2688.

[67] Wang, H.; Wang, X.; Wang, L.; Wang, J.; Jiang, D.; Li, G.; Zhang, Y.; Zhong, H.; Jiang, Y. Phase Transition Mechanism and Electrochemical Properties of Nanocrystalline $MoSe_2$ as Anode Materials for the High Performance Lithium-Ion Battery. *J. Phys. Chem. C* **2015**, *119*, 10197-10205.

[68] Li, X.; Sun, M.; Cheng, S.; Ren, X.; Zang, J.; Xu, T.; Wei, X.; Li, S;; Chen, Q.; Shan, C. Crystallographic-Orientation Dependent Li Ion Migration and Reactions in Layered $MoSe_2$. *2D Mater.* **2019**, *6*, 035027.

[69] Li, H.; Tsai, C.; Koh, A. L.; Cai, L.; Contryman, A. W.; Fragapane, A. H.; Zhao, J.; Han, H. S.; Manoharan, H. C.; Abild-Pedersen, F.; Nørskov, J. K.; Zheng, X. Activating and Optimizing $MoS_2$ Basal Planes for Hydrogen Evolution through the Formation of Strained Sulphur Vacancies. *Nat. Mater.* **2016**, *15*, 48-53.

[70] Duerloo, K. A. N.; Li, Y.; Reed, E. J. Structural Phase Transitions in Two-Dimensional Mo- and W-Dichalcogenide Monolayers. *Nat. Ccommun.* **2014**, *5*, 4214.

[71] Lee, J.; Kim, C.; Choi, K.; Seo, J.; Choi, Y.; Choi, W.; Kim, Y. -M.; Jeong, H.; Lee, J. H.; Kim, G.; Park, H. In-situ Coalesced Vacancies on $MoSe_2$ Mimicking Noble Metal: Unprecedented Tafel Reaction in Hydrogen Evolution. *Nano Energy* **2019**, *63*, 103846.

[72] Tan, C.; Luo, Z.; Chaturvedi, A.; Cai, Y.; Du, Y.; Gong, Y.; Huang, Y.; Lai, Z.; Zhang, X.; Zheng, L.; Qi, X.; Goh, M. H.; Wang, J.; Han, S.; Wu, X. -J.; Gu, L.; Kloc, C.; Zhang, H. Preparation of High-Percentage 1T-Phase Transition Metal Dichalcogenide Nanodots for Electrochemical Hydrogen Evolution. *Adv. Mater.* **2018**, *30*, 1705509.

[73] Cai, L.; Cheng, W.; Yao, T.; Huang, Y.; Tang, F.; Liu, Q.; Liu, W.; Sun, Z.; Hu, F.; Jiang, Y.; Yan, W; Wei, S. High Content Metallic 1T Phase in $MoS_2$-Based Electrocatalyst for Efficient Hydrogen Evolution. *J. Phys. Chem. C* **2017**, *121*, 15071-15077.

[74] He, Y.; Boubeche, M.; Zhou, Y.; Yan, D.; Zeng, L.; Wang, X.; Yan, K.; Luo, H. Topologically Nontrivial 1T'-MoTe$_2$ as Highly Efficient Hydrogen Evolution Electrocatalyst. *J. Phys. Mater.* **2020**, *4*, 014001.

[75] Chen, M.; Zhu, L.; Chen, Q.; Miao, N.; Si, C.; Zhou, J.; Sun, Z. Quantifying the Composition Dependency of the Ground-State Structure, Electronic Property and Phase-Transition Dynamics in Ternary Transition-Metal-Dichalcogenide Monolayers. *J. Mater. Chem. C* **2020**, *8*, 721-733.

[76] Voiry, D.; Salehi, M.; Silva, R.; Fujita, T.; Chen, M.; Asefa, T.; Shenoy, V. B.; Eda, G.; Chhowalla, M. Conducting MoS$_2$ Nanosheets as Catalysts for Hydrogen Evolution Reaction. *Nano Lett.* **2013**, *13*, 6222-6227.

[77] Xie, J.; Qi, J.; Lei, F.; Xie, Y. Modulation of Electronic Structures in Two-Dimensional Electrocatalysts for the Hydrogen Evolution Reaction. *Chem. Commun.* **2020**, *56*, 11910-11930.

[78] Xie, J.; Liu, W.; Zhang, X.; Guo, Y.; Gao, L.; Lei, F.; Tang, B.; Xie, Y. Constructing Hierarchical Wire-On-Sheet Nanoarrays in Phase-Regulated Cerium-Doped Nickel Hydroxide for Promoted Urea Electro-Oxidation. *ACS Mater. Lett.* **2019**, *1*, 103-110.

[79] Hu, X.; Zhang, Q.; Yu, S. Theoretical Insight into the Hydrogen Adsorption on MoS$_2$ (MoSe$_2$) Monolayer as a Function of Biaxial Strain/External Electric Field. *Appl. Surf. Sci.* **2019**, *478*, 857-865.

[80] Yin, Y.; Han, J.; Zhang, Y.; Zhang, X.; Xu, P.; Yuan, Q.; Samad, L.; Wang, X.; Wang, Y.; Zhang, Z.; Zhang, P.; Cao, X.; Song, B.; Jin, S. Contributions of Phase, Sulfur Vacancies, and Edges to the Hydrogen Evolution Reaction Catalytic Activity of Porous Molybdenum Disulfide Nanosheets. *J. Am. Chem. Soc.* **2016**, *13*, 7965-7972.

[81] Joo, J.; Kim, T.; Lee, J.; Choi, S. I.; Lee, K. Morphology-Controlled Metal Sulfides and Phosphides for Electrochemical Water Splitting. *Adv. Mater.* **2019**, *31*, 1806682.

[82] Staszak-Jirkovský, J.; Malliakas, C. D.; Lopes, P. P.; Danilovic, N.; Kota, S. S.; Chang, K. -C.; Genorio, B.; Strmcnik, D.; Stamenkovic, V. R.; Kanatzidis, M. G.; Markovic, N. M. Design of Active and Stable Co-Mo-S$_x$ Chalcogels as pH-Universal Catalysts for the Hydrogen Evolution Reaction. *Nat. Mater.* **2016**, *15*, 197-203.

[83] Ge, Y.; Gao, S. -P.; Dong, P.; Baines, R.; Ajayan, P. M.; Ye, M.; Shen, J. Insight into the Hydrogen Evolution Reaction of Nickel Dichalcogenide Nanosheets: Activities Related to Non-Metal Ligands. *Nanoscale* **2017**, *9*, 5538-5544.

[84] Rasmussen, F. A.; Thygesen, K. S. Computational 2D Materials Database: Electronic Structure of Transition-Metal Dichalcogenides and Oxides. *J. Phys. Chem. C* **2015**, *119*, 13169-13183.

[85] De Silva, U.; Masud, J.; Zhang, N.; Hong, Y.; Liyanage, W. P. R.; Asle Zaeem, M.; Nath, M. Nickel Telluride as a Bifunctional Electrocatalyst for Efficient Water Splitting in Alkaline Medium. *J. Mater. Chem.* **2018**, *6*, 7608-7822.

[86] Kosmala, T.; Coy Diaz, H.; Komsa, H. P.; Ma, Y.; Krasheninnikov, A. V.; Batzill, M.; Agnoli, S. Metallic Twin Boundaries Boost the Hydrogen Evolution Reaction on the Basal Plane of Molybdenum Selenotellurides. *Adv. Energy Mater.* **2018**, *8*, 1800031.

[87] Seok, J.; Lee, J. H.; Bae, D.; Ji, B.; Son, Y. W.; Lee, Y. H.; Yang, H.; Cho, S. Hybrid Catalyst with Monoclinic MoTe$_2$ and Platinum for Efficient Hydrogen Evolution. *APL Mater.* **2019**, *7*, 071118.

[88] Yoo, D.; Kim, M.; Jeong, S.; Han, J.; Cheon, J. Chemical Synthetic Strategy for Single-Layer Transition-Metal Chalcogenides. *J. Am. Chem. Soc*. **2014**, *136*, 14670-14673.

[89] Balasingam, S. K.; Lee, J. S.; Jun, Y. Few-Layered MoSe$_2$ Nanosheets as an Advanced Electrode Material for Supercapacitors. *Dalton Trans.* **2015**, *44*, 15491-15498.

[90] Jiang, M.; Zhang, J.; Wu, M.; Jian, W.; Xue, H.; Ng, T. -W.; Lee, C. -S.; Xu, J. Synthesis of 1T-MoSe$_2$ Ultrathin Nanosheets with an Expanded Interlayer Spacing of 1.17 nm for Efficient Hydrogen Evolution Reaction. *J. Mater. Chem. A* **2016**, *4*, 14949-14953.

[91] Zhou, R.; Wang, H.; Chang, J.; Yu, C.; Dai, H.; Chen, Q.; Zhou, J.; Yu, H.; Sun, G.; Huang, W. Ammonium Intercalation Induced Expanded 1T-Rich Molybdenum Diselenides for Improved Lithium Ion Storage. *ACS Appl. Mater. Interfaces* **2021**, *13*, 17459-17466.

[92] Yi, Y.; Sun, Z.; Li, C.; Tian, Z.; Lu, C.; Shao, Y.; Li, J.; Sun, J.; Liu, Z. Designing 3D Biomorphic Nitrogen-Doped $MoSe_2$/Graphene Composites toward High-Performance Potassium-Ion Capacitors. *Adv. Funct. Mater.* **2020**, *30*, 1903878.

[93] Bhat, K. S.; Nagaraja, H. S. Performance Evaluation of Molybdenum Dichalcogenide ($MoX_2$; X= S, Se, Te) Nanostructures for Hydrogen Evolution Reaction. *Int. J. Hydrog. Energy* **2019**, *44*, 17878-17886.

[94] Hu, C.; Zhang, L.; Zhao, Z. J.; Luo, J.; Shi, J.; Huang, Z.; Gong, J. Edge Sites with Unsaturated Coordination on Core-Shell $Mn_3O_4@Mn_xCo_{3x}O_4$ Nanostructures for Electrocatalytic Water Oxidation. *Adv. Mater.* **2017**, *29*, 1701820.

[95] Chia, X.; Pumera, M. Inverse Opal-Like Porous $MoSe_x$ Films for Hydrogen Evolution Catalysis: Overpotential-Pore Size Dependence. *ACS Appl. Mater. Interfaces* **2018**, *10*, 4937-4945.

[96] Bhat, K. S.; Barshilia, H. C.; Nagaraja, H. S. Porous Nickel Telluride Nanostructures as Bifunctional Electrocatalyst towards Hydrogen and Oxygen Evolution Reaction. *Int. J. Hydrog. Energy* **2017**, *42*, 24645-24655.

[97] Ndala, Z.; Shumbula, N.; Nkabinde, S.; Kolokoto, T.; Nchoe, O.; Shumbula, P.; Tetana, Z. N.; Linganiso, E. C.; Gqoba, S. S.; Moloto, N. Evaluating the Effect of Varying the Metal Precursor in the Colloidal Synthesis of $MoSe_2$ Nanomaterials and Their Application as Electrodes in the Hydrogen Evolution Reaction. *Nanomaterials* **2020**, *10*, 1786.

[98] Kwon, I. S.; Kwak, I. H.; Debela, T. T.; Abbas, H. G.; Park, Y. C.; Ahn, J. P.; Park, J.; Kang, H. S. Se-Rich $MoSe_2$ Nanosheets and Their Superior Electrocatalytic Performance for Hydrogen Evolution Reaction. *ACS Nano* **2020**, *14*, 6295-6304.

[99] Najafi, L.; Bellani, S.; Oropesa-Nunez, R.; Ansaldo, A.; Prato, M.; Del Rio Castillo, A. E.; Bonaccorso, F. Engineered $MoSe_2$-Based Heterostructures for Efficient Electrochemical Hydrogen Evolution Reaction. *Adv. Energy Mater.* **2018**, *8*, 1703212.

[100] Xia, B.; Wang, T.; Jiang, X.; Zhang, T.; Li, J.; Xiao, W.; Xi, P.; Gao, D.; Xue, D.; Ding, J. $Ar^{2+}$ Beam Irradiation-Induced Multivancancies in $MoSe_2$ Nanosheet for Enhanced Electrochemical Hydrogen Evolution. *ACS Energy Lett.* **2018**, *3*, 2167-2172.

[101] Tang, Q. Tuning the Phase Stability of Mo-based TMD Monolayers through Coupled Vacancy Defects and Lattice Strain. *J. Mater. Chem. C* **2018**, *6*, 9561-9568.

[102] Yin, Y.; Zhang, Y.; Gao, T.; Yao, T.; Zhang, X.; Han, J.; Wang, X.; Zhang, Z.; Xu, P.; Zhang, P.; Cao, X.; Song, B.; Jin, S. Synergistic Phase and Disorder Engineering in 1T-$MoSe_2$ Nanosheets for Enhanced Hydrogen-Evolution Reaction. *Adv. Mater.* **2017**, *29*, 1700311.

[103] Zhang, H.; Yu, L.; Chen, T.; Zhou, W.; Lou, X. W. Surface Modulation of Hierarchical $MoS_2$ Nanosheets by Ni Single Atoms for Enhanced Electrocatalytic Hydrogen Evolution. *Adv. Funct. Mater.* **2018**, *28*, 1807086.

[104] Yu, C.; Cao, Z.; Chen, S.; Wang, S.; Zhong, H. Promoting the Hydrogen Evolution Performance of 1T-$MoSe_2$-Se: Optimizing the Two-Dimensional Structure of $MoSe_2$ by Layered Double Hydroxide Limited Growth. *Appl. Surf. Sci.* **2020**, *509*, 145364.

[105] Yi, J.; Li, H.; Gong, Y.; She, X.; Song, Y.; Xu, Y.; Deng, J.; Yuan, S.; Xu, H.; Li, H. Phase and Interlayer Effect of Transition Metal Dichalcogenide Cocatalyst toward Photocatalytic Hydrogen Evolution: the Case of $MoSe_2$. *Appl. Catal. B-Environ.* **2019**, *243*, 330-336.

[106] Zhang, Y.; Gong, Q.; Li, L.; Yang, H.; Li, Y.; Wang, Q. $MoSe_2$ Porous Microspheres Comprising Monolayer Flakes with High Electrocatalytic Activity. *Nano Res.* **2015**, *8*, 1108-1115.

[107] Duan, X.; Wang, C.; Pan, A.; Yu, R.; Duan, X. Two-Dimensional Transition Metal Dichalcogenides as Atomically Thin Semiconductors: Opportunities and Challenges. *Chem. Soc. Rev.* **2015**, *44*, 8859-8876.

[108] Takahashi, Y.; Nakayasu, Y.; Iwase, K.; Kobayashi, H.; Honma, I. Supercritical Hydrothermal Synthesis of MoS$_2$ Nanosheets with Controllable Layer Number and Phase Structure. *Dalton Trans.* **2020**, *49*, 9377-9384.

[109] Zhang, J.; Wang, T.; Liu, P.; Liu, Y.; Ma, J.; Gao, D. Enhanced Catalytic Activities of Metal-Phase-Assisted 1T@2H-MoSe$_2$ Nanosheets for Hydrogen Evolution. *Electrochim. Acta* **2016**, *217*, 181-186.